\documentclass[prd,reprint,amsmath,amssymb,preprintnumbers,superscriptaddress,nofootinbib]{revtex4-1}

\usepackage{subfigure}
\usepackage{enumitem}
\usepackage{color}
\usepackage{graphicx}
\usepackage{amsmath}
\usepackage{hyperref}

\def\beq{\begin{equation}}
\def\eeq{\end{equation}}
\def\bea{\begin{eqnarray}} 
\def\eea{\end{eqnarray}}
\def\nn{\nonumber}
\def\gev{\rm GeV}

\def\mev{{\rm MeV}}

\def\eps{\varepsilon}

\newcommand{\lsim}{
\mathrel{\hbox{\rlap{\hbox{\lower4pt\hbox{$\sim$}}}\hbox{$<$}}}}
\newcommand{\gsim}{
\mathrel{\hbox{\rlap{\hbox{\lower4pt\hbox{$\sim$}}}\hbox{$>$}}}}

\begin{document}
    
\preprint{CTPU-PTC-18-16}
\title{\boldmath Implications of the dark axion portal for the muon $g-2$, $B$-factories, \\ fixed target neutrino experiments and beam dumps}

\author{Patrick deNiverville}
\affiliation{Center for Theoretical Physics of the Universe, IBS, Daejeon 34126, Korea}
\author{Hye-Sung Lee}
\affiliation{Department of Physics, KAIST, Daejeon 34141, Korea}
\author{Min-Seok Seo}
\affiliation{Department of Physics, Chung-Ang University, Seoul 06974, Korea}
\date{June 2018}
\begin{abstract}
The dark axion portal is a recently introduced portal between the standard model and the dark sector.
It connects both the dark photon and the axion (or axion-like particle) to the photon simultaneously through an anomaly triangle.
While the vector portal and the axion portal have been popular venues to search for the dark photon and axion, respectively, the new portal provides new detection channels if they coexist.
The dark axion portal is not a result of the simple combination of the two portals, and its value is not determined by the other portal values; it should be tested independently.
In this paper, we discuss implications of the new portal for the leptonic $g-2$, $B$-factories, fixed target neutrino experiments and beam dumps.
We provide the model-independent constraints on the axion-photon-dark photon coupling and discuss the sensitivities of the recently activated Belle-II experiment, which will play an important role in testing the new portal.
\end{abstract}
\maketitle

\section{Introduction}
\label{sec:introduction}
Our universe can be divided into two sectors: the visible and the dark.
The visible sector of the universe is comprised of the standard model (SM) particles, whose constituents were all identified with the last discovery of the Higgs boson in 2012 \cite{Aad:2012tfa,Chatrchyan:2012xdj}.
Just as the SM has various kinds of fermions, gauge bosons, and a scalar boson, the dark sector might also have a rich spectrum of dark fermions, dark gauge bosons, and dark scalar bosons rather than a single dark matter (DM) particle.
Although there has been no discovery of the dark sector particles so far, the existence of the dark sector is backed by significant observational evidence of DM \cite{Patrignani:2016xqp}.

It is natural to expect that there is a connection between the SM particles and DM particles other than gravity as the typical explanations (freeze-out, freeze-in \cite{Hall:2009bx}) of the dark matter relic density require an interaction between the two sectors.
While it is possible the dark sector particles carry the SM weak charges (as in many supersymmetric dark matter models), it may also be very possible they do not carry any charges under the SM gauge symmetries.
Even in the latter case, two separate sectors might still be able to communicate with each other if there is a `portal,' a way to connect the visible sector particles and the dark sector particles through a mixing or a loop-effect.

There have been four popular portals: \\
$~~~~~~~~~~$ (i) Vector portal:  $\frac{\varepsilon}{2 \cos\theta_W} B_{\mu\nu} Z'^{\mu\nu}$, \\
$~~~~~~~~~~~$(ii) Axion portal:  $\frac{G_{a\gamma\gamma}}{4} a F_{\mu\nu} \tilde F^{\mu\nu} , \cdots$, \\
$~~~~~~~~~~~$(iii) Higgs portal: $\kappa |S|^2 H^\dagger H , \cdots$, \\
$~~~~~~~~~~~$(iv) Neutrino portal: $y_N LHN$. \\
The constraints on these portals can be found in Refs.~\cite{Essig:2013lka,Patrignani:2016xqp,Alexander:2016aln}.
The relic DM can be either a portal particle or a particle coupled to portal particles via hidden interactions (for examples see Refs.~\cite{ArkaniHamed:2008qn,Nomura:2008ru}).

The vector portal \cite{Holdom:1985ag} is a mixing between a SM gauge boson and a dark sector gauge boson (such as dark photon \cite{ArkaniHamed:2008qn} and the dark $Z$ \cite{Davoudiasl:2012ag}, which is a variant of the dark photon with an axial coupling \cite{Davoudiasl:2012qa,Davoudiasl:2013aya,Lee:2013fda,Davoudiasl:2014kua,Kong:2014jwa,Kim:2014ana,Davoudiasl:2015bua}).
The axion portal connects the axion or axion-like particle to a pair of the SM gauge bosons. Recently, it was pointed out that a `dark axion portal' \cite{Kaneta:2016wvf}) may exist, connecting the dark photon and axion to the SM.
The new portal is independent from the vector and axion portals as it arises from a different mechanism.

When a new portal is introduced, it can provide new opportunities to search for dark sector particles \cite{Essig:2013lka}.
For some of the recent studies using the dark axion portal, see Refs.~\cite{Choi:2016kke,Kaneta:2017wfh,Agrawal:2017eqm,Kitajima:2017peg,Choi:2018dqr}.
Because of the very small couplings between the dark sector and the SM particles, their masses can be much smaller than the typical (electroweak - TeV) scale of new physics.
As a matter of fact, most of the studies of the portal focus on the rather light masses as we can see in the dark photon and axion (or axion-like particle) cases.
(For some mechanisms to introduce very light particles, see Refs.~\cite{Kim:1979if,Shifman:1979if,Lee:2016ejx}.)
Various studies can be summarized in a similar fashion as in Ref.~\cite{Jaeckel:2013ija} that show the constraints on the vast parameter space of the portal particle (mass and coupling).

In this paper, we study the implications of the dark axion portal for a roughly MeV - 10 GeV scale dark photon, the mass range focused on by the typical intensity frontier new physics \cite{Essig:2013lka}.
We investigate possible signals of the new portal at the $B$-factories and use the existing data from the BaBar experiment to constrain the axion-photon-dark photon coupling.
We also study the sensitivities of the new Belle-II experiment for both Phase II and Phase III running.
Belle-II began data taking in April 2018 with a partially complete detector, and will be one of the major players in the intensity frontier  physics over the next decade.
We also study the implications for the muon and electron $g-2$ and determine the constraints on the new coupling from existing measurements.
Finally, we study possible dark axion portal signals at the LSND and MiniBooNE fixed target neutrino experiments, and the CHARM proton beam dump.

While we focus on MeV - 10 GeV scale physics, much heavier particles can be searched for by energy frontier experiments such as the LHC experiments; much lighter ones may be observed by the cosmic frontier observations such as stellar cooling and supernovae.

The exact nature of the axion-photon-dark photon coupling is model dependent and the predictions may change depending on other related couplings (such as the axion-photon-photon and axion-dark photon-dark photon couplings), but we will treat it in a model-independent way by taking the limit where only axion-photon-dark photon coupling is relevant.

In Sec.~\ref{sec:DAP}, we briefly discuss the dark axion portal vertex, and elaborate on our parameterization.
In Sec.~\ref{sec:calc}, we discuss the search channels and constraints for the dark axion portal from the BaBar and Belle-II experiments.
In Sec.~\ref{ssec:g_2}, we discuss the contributions to the electron and muon $g-2$ from the new axion-photon-dark photon vertex and obtain constraints from current measurements.
In Sec.~\ref{sec:ftnf}, we study the ability of the LSND and MiniBooNE fixed target neutrino facilities to constrain the dark axion portal.
In Sec.~\ref{sec:beamdump}, we place limits on the dark axion portal parameter space with two analyses at the CHARM experiment, and make a few comments on electron beam dumps.
In Sec.~\ref{sec:summary}, we provide a summary of our study and future directions.

\section{Dark axion portal}
\label{sec:DAP}
The axion portal and the dark axion portal terms \cite{Kaneta:2016wvf} can be written as following.
\begin{eqnarray}
&& {\cal L}_\text{axion portal} = \frac{G_{agg}}{4} a G_{\mu\nu}\tilde G^{\mu\nu} + \frac{G_{a\gamma\gamma}}{4} a F_{\mu\nu}\tilde F^{\mu\nu} + \cdots ~~ \\
&&{\cal L}_\text{dark axion portal} = \frac{G_{a\gamma^\prime\gamma^\prime}}{4}  a Z'_{\mu\nu}\tilde Z'^{\mu\nu} +\frac{G_{a\gamma\gamma^\prime}}{2} a F_{\mu\nu}\tilde Z'^{\mu\nu} ~~~ 
\end{eqnarray}

The dark axion and the axion portal are constructed using the anomaly triangle and the actual couplings depend on the details of the model.
For instance, in the dark KSVZ axion model introduced in Ref.~\cite{Kaneta:2016wvf}, the portal couplings are given as
\bea
G_{a\gamma\gamma} &=& \frac{e^2}{8\pi^2} \frac{PQ_\Phi}{f_a} \Big[ 2 N_C Q_\psi^2 - \frac{2}{3} \frac{4+z}{1+z} \Big] , \label{eq:Gagg} \\
G_{a\gamma\gamma'} &\simeq& \frac{e e'}{8\pi^2} \frac{PQ_\Phi}{f_a} \big[ 2 N_C D_\psi Q_\psi \big] + \eps G_{a\gamma\gamma} , \\
G_{a\gamma'\gamma'} &\simeq& \frac{e'^2}{8\pi^2} \frac{PQ_\Phi}{f_a} \big[ 2 N_C D_\psi^2 \big] + 2 \eps G_{a\gamma\gamma'},
\eea
where $N_C = 3$ is the color factor.
$e$ ($e'$) and $Q_\psi$ ($D_\psi$) are the electric (dark) coupling constant and charge of the exotic quarks in the anomaly triangle.
$f_a / PQ_\Phi$ is the mass scale of the exotic quarks.
$z = m_u / m_d \simeq 0.56$ is the mass ratio of the $u$ and $d$ quarks.
$\varepsilon$ is the vector portal coupling which we take to be 0 in our study.

While one can consider the coupling in the context of a specific model that can decide the couplings in terms of the model parameters and provide connections among them, we focus on the limit where a model independent treatment makes sense and consider only the $G_{a\gamma\gamma'}$ coupling. Specifically, we take $m_a \ll m_{\gamma'}$ and also take the view the model-specific part of the $G_{a\gamma\gamma}$ such as the electric charge contribution are arranged to make $G_{a\gamma\gamma}$ small enough to neglect its effect in the analysis we perform in this paper.

 We do not claim that the $a$ should be the QCD axion, but take it to an axion-like particle with a mass much smaller than that of the $\gamma^\prime$. The $G_{a\gamma'\gamma'}$ is nonzero, but the on-shell decay process $a \to \gamma' \gamma'$ is kinematically forbidden, while the off-shell process would be negligibly small.
While the decay $a \to \gamma \gamma$ is allowed, by considering a very small $m_a$, the aforementioned arrangement to minimize $G_{a\gamma\gamma}$ would ensure the the $a$ is sufficiently long-lived to escape the $B$-factory detectors before its decay and its effect on the lepton $g-2$ is suppressed, making the effect of the $G_{a\gamma\gamma}$ negligible in our analysis.
More general cases and their implications will be studied in subsequent works.

\section{B Factories}
\label{sec:calc}

\subsection{BaBar}
\label{ssec:visible}
BaBar \cite{Aubert:2001tu} is an asymmetric electron-positron collider with a 9\,GeV electron beam and a 3.1\,GeV positron beam for a center of mass energy of 10.5\,GeV. The experiment collected an integrated luminosity of over 500\,fb$^{-1}$ \cite{Lees:2013rw} between 1999 and 2008, but the monophoton trigger was only implemented for its final running period.

\begin{figure}[t]
 \centerline{\includegraphics[width=0.3\textwidth]{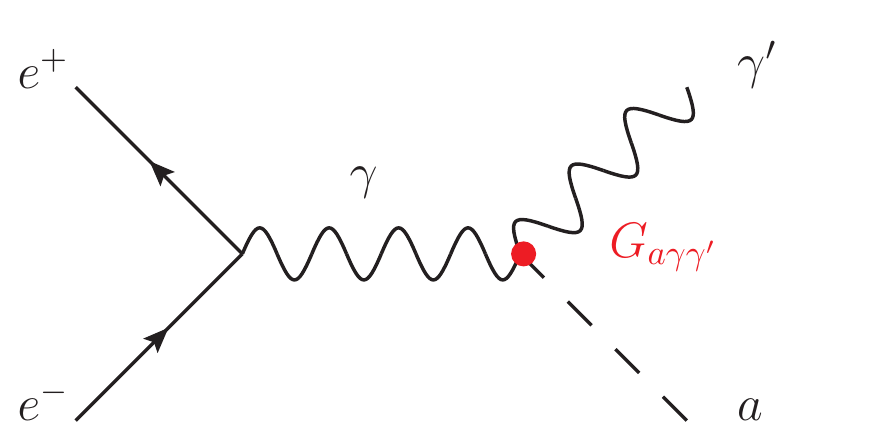}}
 \caption{Electron-positron annihilation to on-shell $a$ and $\gamma^\prime$. Observable at $B$ factories as a monophoton produced through subsequent decay $\gamma^\prime \to \gamma a$.}
 \label{fig:feyn_vis}
\end{figure}

\begin{table}[b]
 \centering
 \begin{tabular}{lcc}
 \hline
 \hline
 & Low-Cut & High-Cut\\
 \hline
 $E_\gamma^*$ & $[2.2,3.7]$GeV & $[3.2, 5.5]$GeV \\
 \hline
 $\cos \theta_\gamma^*$ & $[-0.46,0.46]$ & $[-0.31,0.6]$ \\
 \hline
 Luminosity & 19 $\mathrm{fb}^{-1}$ & 28 $\mathrm{fb}^{-1}$\\
 \hline
 Efficiency & 55\% & 30\% \\
 \hline
 \end{tabular}
\caption{Kinematic cuts on the BaBar Low-Cut and High-Cut samples. $E_\gamma^*$ is the center of mass energy of the detected photon and $\theta_\gamma^*$ is the angle of the photon relative to the beam axis in the center of mass frame.}
\label{tab:params}
\end{table}

\begin{figure}[t]
 \centerline{\includegraphics[height=0.33\textwidth]{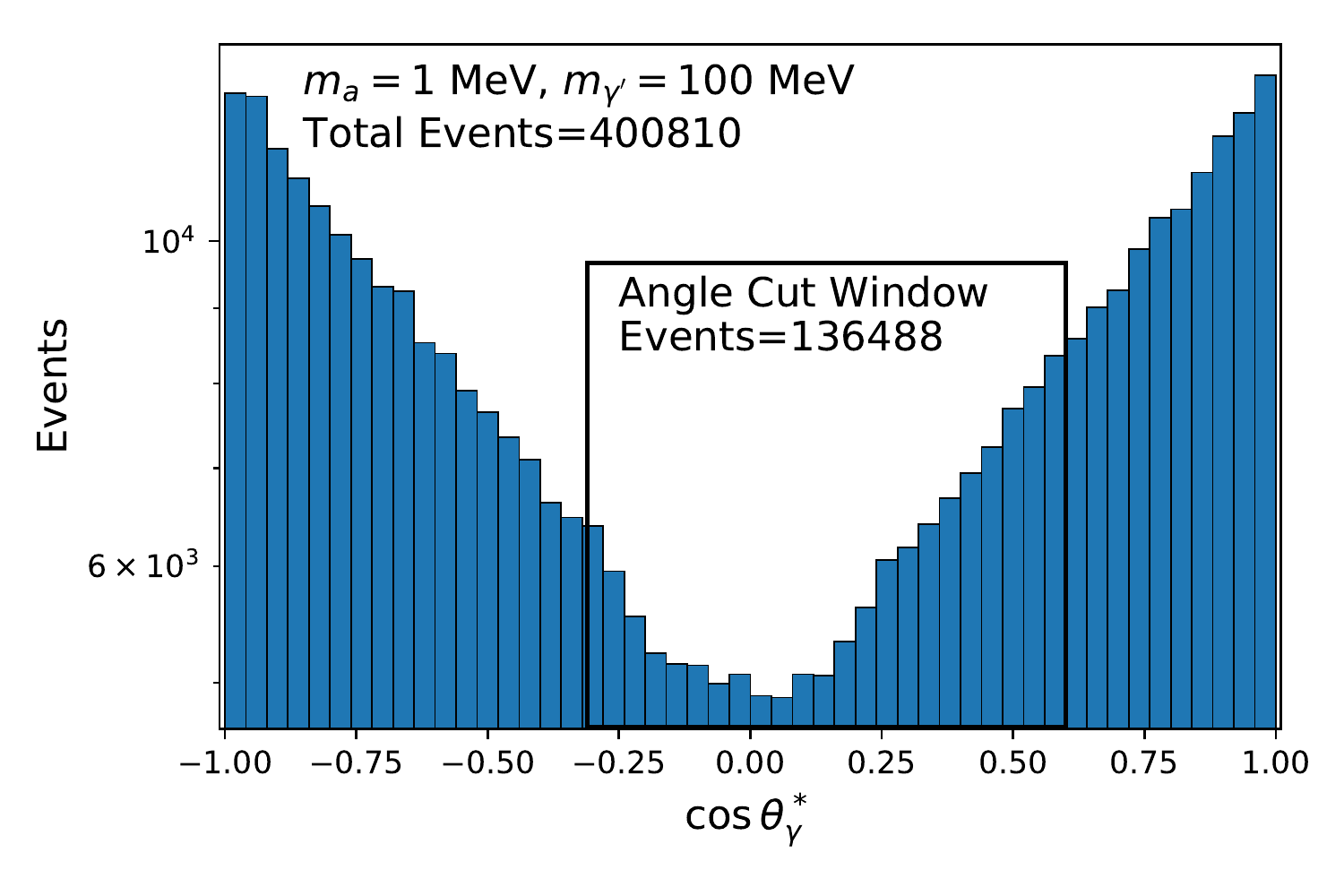}}
 \caption{As an illustration, we provide a histogram of the cosine of the center-of-mass emission angle $\cos \theta_\gamma^*$ for a sample of $4\times10^5$ secondary photons produced through $e^+ e^- \to a  (\gamma^\prime \to a \gamma)$ for High-Cut energies. The boxed window reflects the imposed High-Cut $\cos \theta_\gamma^*$ range detailed in Table \ref{tab:params}. Of the $10^6$ events initially generated, $1.36\times10^5$ survive both the energy and angle cuts. The bin size for this histogram was chosen to provide a good representation of the angular distribution, but is not relevant to the analysis.}
 \label{fig:angle_hist_2}
\end{figure}

We examine the process $e^+ e^- \to a \gamma^\prime$ shown in Fig. \ref{fig:feyn_vis}, and calculated with FeynCalc \cite{Shtabovenko:2016sxi,Mertig:1990an} to be
\begin{equation}
 \frac{d\sigma}{dt} = \frac{\alpha_\text{EM} G_{a\gamma\gamma^\prime}^2}{16 s^3}\left(2 m_{\gamma^\prime}^4 - 2m_{\gamma^\prime}^2(s+t+u)+t^2+u^2\right),
\label{eq:ee_ann}
\end{equation}
where $s=(p_{e^-}+p_{e^-})^2$, $t=(p_{\gamma^\prime}-p_{e^-})^2$ and $u=(p_{a}-p_{e^-})^2$ are the Mandelstam variables. This process can result in the production of a monophoton final state through a subsequent $\gamma^\prime \to a\gamma$ decay, so long as the $\gamma^\prime$ is reasonably prompt. We will assume that $m_a \ll m_{\gamma^\prime}$, and that the $a$ is sufficiently long-lived to escape the BaBar detector before decaying radiatively. We follow the approach of Ref.~\cite{Essig:2013vha} in using BaBar's $\Upsilon(3S) \to \gamma$A0 data, where A0 is some invisibly decaying scalar particle \cite{Aubert:2008as}\footnote{The limits we will place using this data could potentially be improved by using a larger BaBar analysis that included a background model \cite{Lees:2017lec}.}. This set of data records the measured center-of-mass energy of detected monophotons, $E_\gamma^\star$. The data is divided into overlapping Low-Cut and High-Cut $E_\gamma^\star$ domains, where the Low-Cut domain is $E_\gamma^\star\in[2.2,3.7]$\,GeV, and the High-Cut domain is $E_\gamma^\star\in[3.2,5.5]$\,GeV. See Table \ref{tab:params} for a summary of the cuts, luminosities and efficiencies.  

Samples of $10^6$ $e^+ e^- \to a\gamma^\prime $ events were generated with CalcHEP 3.6.27 \cite{Pukhov:1999gg,Belyaev:2012qa} for 45 dark photon masses.  The subsequent $\gamma^\prime \to a\gamma$ decays were simulated using an external Python code. As in Ref. \cite{Essig:2013vha}, the simulated photons were smeared using a Crystal Ball function (see Ref.~\cite{Skwarnicki:1986xj}) with $n=1.79$, $\alpha=0.811$ and $\sigma/(E_\gamma^\star) = 0.015\times \left(\gev/E_\gamma^\star\right)^{3/4} + 0.01$. In the absence of a background model, we will place a conservative limit on the coupling constant $G_{a \gamma \gamma^\prime}$ by treating all measured events as signal, and taking the maximum value of $G_{a \gamma \gamma^\prime}$ for which the theory prediction does not exceed the measured number of events in any bin of either the High- or Low-Cut data by more than $2\sigma$.

\begin{figure}[t]
 \centerline{\includegraphics[width=0.48\textwidth]{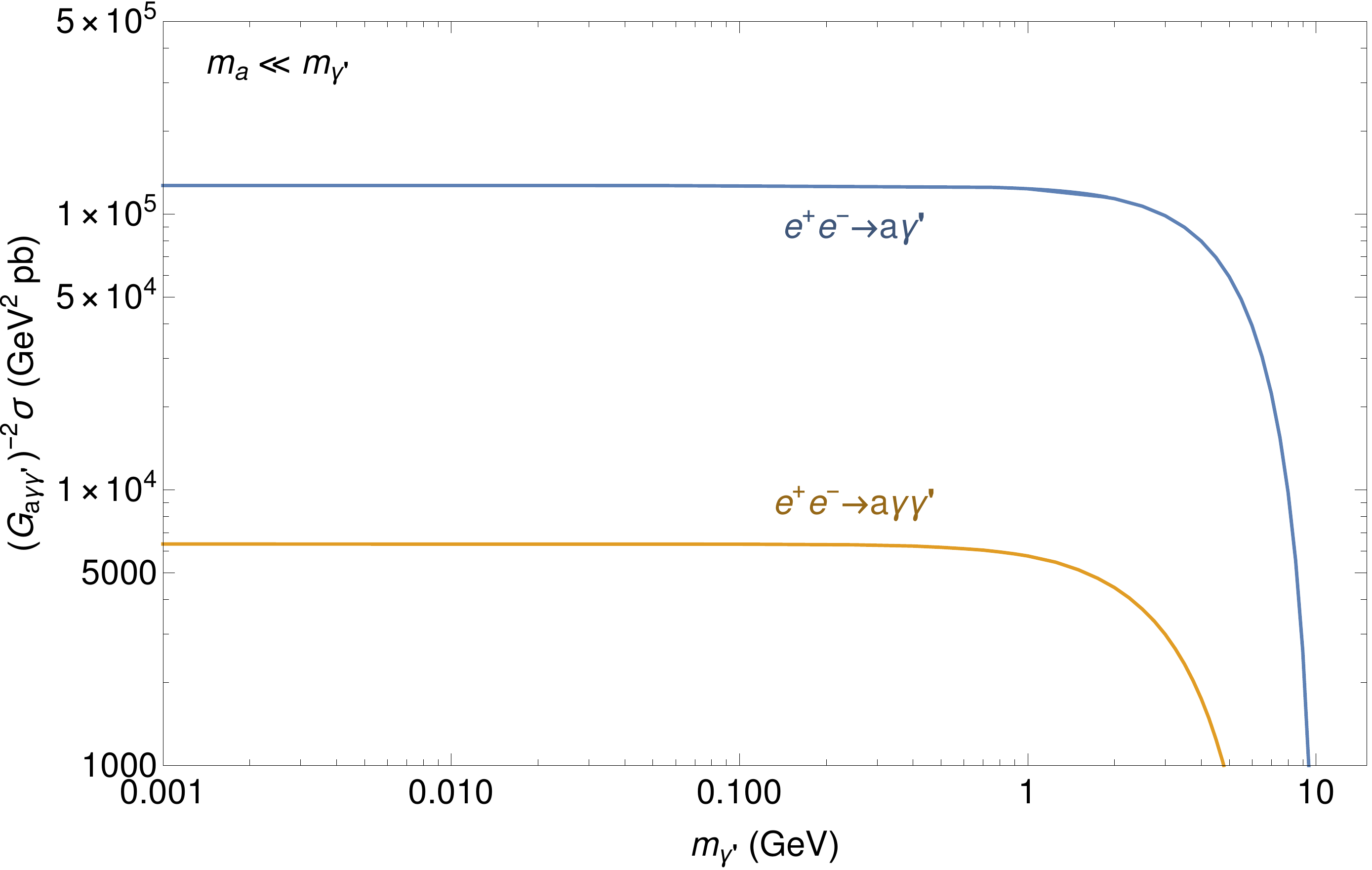}}
 \caption{Cross sections for the processes $e^+ e^- \to a \gamma^\prime$ and $e^+ e^- \to a \gamma \gamma^\prime$ with $m_a \ll m_{\gamma^\prime}$ $E_\gamma \in [2.2,5.5]\,\gev$.
The cross section of $e^+ e^- \to a \gamma \gamma^\prime$ is heavily suppressed relative to $e^+ e^- \to a \gamma^\prime$ due in part to the three body final state.}
 \label{fig:cross_section_2}
\end{figure}

A sample of the angular distribution of photons produced in the chain $e^+e^-\to a(\gamma^\prime \to a\gamma)$ is shown in Fig.~\ref{fig:angle_hist_2}.
The events from $e^+ e^- \to a\gamma  (\gamma^\prime \to a \gamma)$ could be potentially relevant, as the primary photon is preferentially emitted along the beam axis while the secondary photon produced through the $\gamma^\prime$ decay has a much broader angular distribution and frequently passes the required monophoton cuts. However, this process possesses a much smaller cross section than pure annihilation (compare the lines shown in Fig.~\ref{fig:cross_section_2}) and would contribute at a subleading level to the observed monophoton signal.

\begin{figure*}[t]
\centerline{\includegraphics[width=0.7\textwidth]{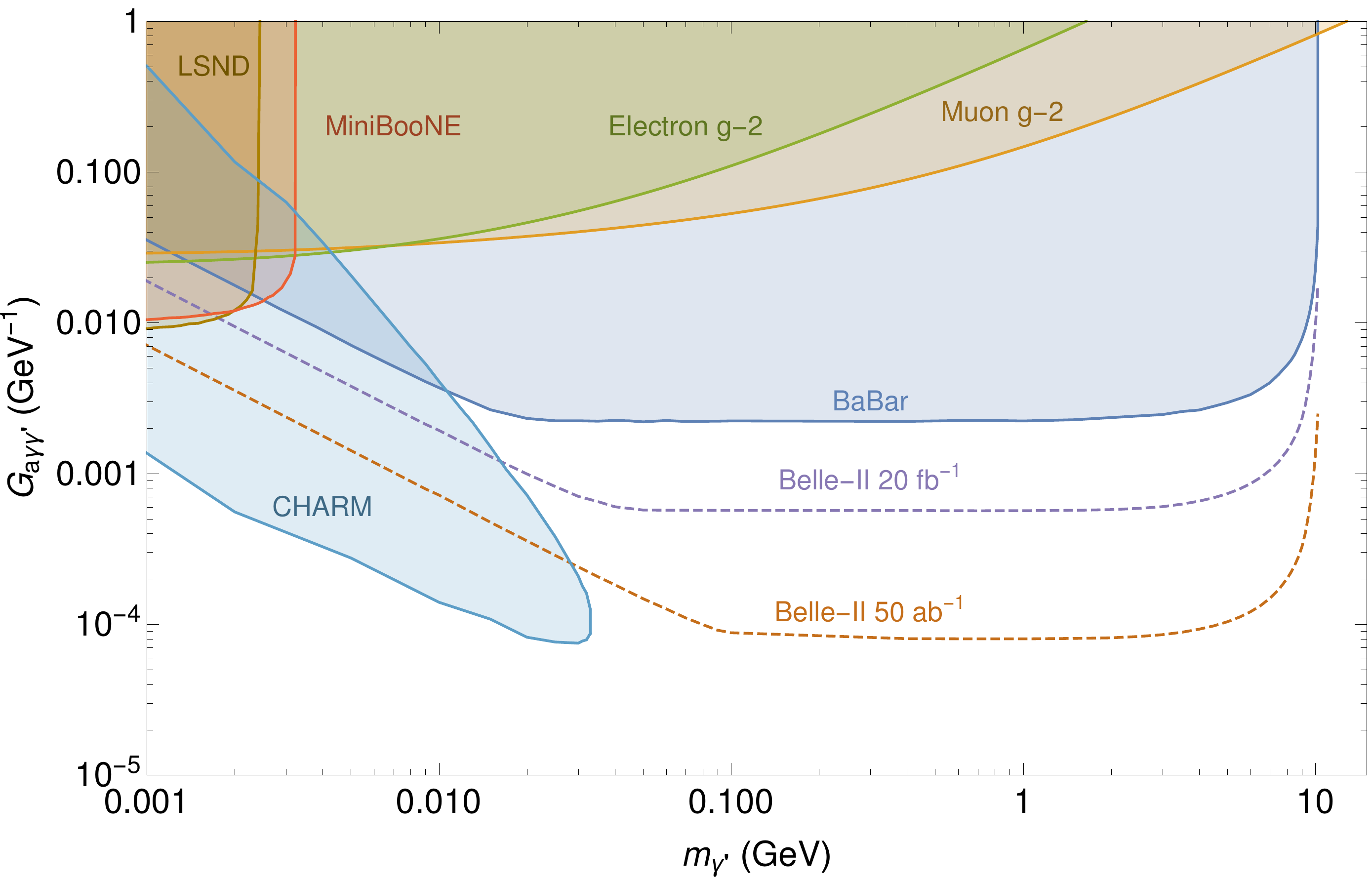}}
  \caption{Limits placed on $G_{a\gamma\gamma^\prime}$ for the hierarchical mass scenario with $m_a\ll m_{\gamma^\prime}$. The BaBar line refers to searches for monophotons produced through $e^+e^-\to a\gamma^\prime$ followed by the decay $\gamma^\prime  \to a\gamma$. Also shown are projected sensitivities for a similar search in Phase-II and III of the Belle-II experiment. The LSND and MiniBooNE lines reflect a search for excess neutral current elastic scattering events in the LSND and MiniBooNE detectors. The CHARM constraint is the result of a search for $\gamma^\prime \to a\gamma$ decays in the CHARM fine-grain detector. The electron and muon g-2 constraints represent parameter space for which the scenario is excluded due to changes in the lepton anomalous magnetic moment that are incompatible with current experimental measurements.}
  \label{fig:limits_2}
\end{figure*}

We show the limits obtained by BaBar, using only $e^+e^-\to a(\gamma^\prime \to a\gamma$), for $m_a \ll m_{\gamma^\prime}$ in Fig.~\ref{fig:limits_2}.
For $m_{\gamma^\prime} \le 100\,\mev$, the lifetime can become sufficiently large for relevant values of $G_{a\gamma\gamma^\prime}$ that dark photons begin to escape the detector before they decay, reducing the number of observed monophoton events:
\begin{align}
 \label{eq:gammap_tau}
 c\tau_{\gamma^\prime} &\approx \frac{5.95 \times 10^{-14} \, \mathrm{m}\cdot \mathrm{GeV}}{G_{a\gamma\gamma^\prime}^2 m_{\gamma^\prime}^3} \quad (\text{for }m_a \ll m_{\gamma^\prime})\\
                       &\approx 60\,\text{m}\times\left(\frac{10^{-3}\,\gev^{-1}}{G_{a\gamma\gamma^\prime}}\right)^2\times\left(\frac{1\,\mev}{m_{\gamma^\prime}}\right)^3 .
\end{align}
This is reflected in a pronounced shoulder in the limit contour as, due to the decline in $m_{\gamma^\prime}$, the lifetime of the $\gamma^\prime$ becomes of $\mathcal{O}(1\,\text{m})$, a length comparable in size to the BaBar detector.

\subsection{Belle-II}
\label{ssec:belle2}
The Belle-II experiment \cite{Abe:2010gxa} is the successor to the Belle and BaBar experiments, and has recently begun taking data as part of Phase II of its operations. Phase II aims to record 20\,fb$^{-1}$ of integrated luminosity with a partially completed detector, while Phase III of the experiment will take 50\,ab$^{-1}$ of data with the completed detector and the SuperKEKB particle accelerator. Unlike BaBar, Belle-II will run with a monophoton trigger for the entirety of its run. To estimate the sensitivity of Belle-II to the $a$-$\gamma$-$\gamma^\prime$ vertex, samples of $10^6$ $e^+ e^- \to \gamma \gamma^\prime a$ events were generated with CalcHEP 3.6.27 \cite{Pukhov:1999gg,Belyaev:2012qa} for 48 values of $m_{\gamma^\prime}$ chosen so as to smoothly render all features of the contour. Photons were generated through the decay of the $\gamma^\prime$ and the number satisfying the preliminary cuts shown in Ref. \cite{HeartyBelleII} were recorded. This preliminary analysis predicted that 300 background events would survive these cuts for 20\,fb$^{-1}$ of data, and we scale this to $7.5\times10^{5}$ background events for 50\,ab$^{-1}$.

We show contours for the expected Phase-II and Phase-III luminosities, with $2\sqrt{300}$ events for the $20\,\text{fb}^{-1}$ contour, and scale up these backgrounds to $2\sqrt{7.5\times10^5}$ events for the $50\,\text{ab}^{-1}$ contour for $m_a \ll m_{\gamma^\prime}$ in Fig. \ref{fig:limits_2}. Thanks to a combination of greater luminosity and generous angular cuts, Belle-II is capable of probing far smaller values of $G_{a \gamma \gamma^\prime}$ than BaBar.

\section{\boldmath Lepton $g-2$}
\label{ssec:g_2}
\begin{figure}[t]
 \centerline{\includegraphics[width=0.3\textwidth]{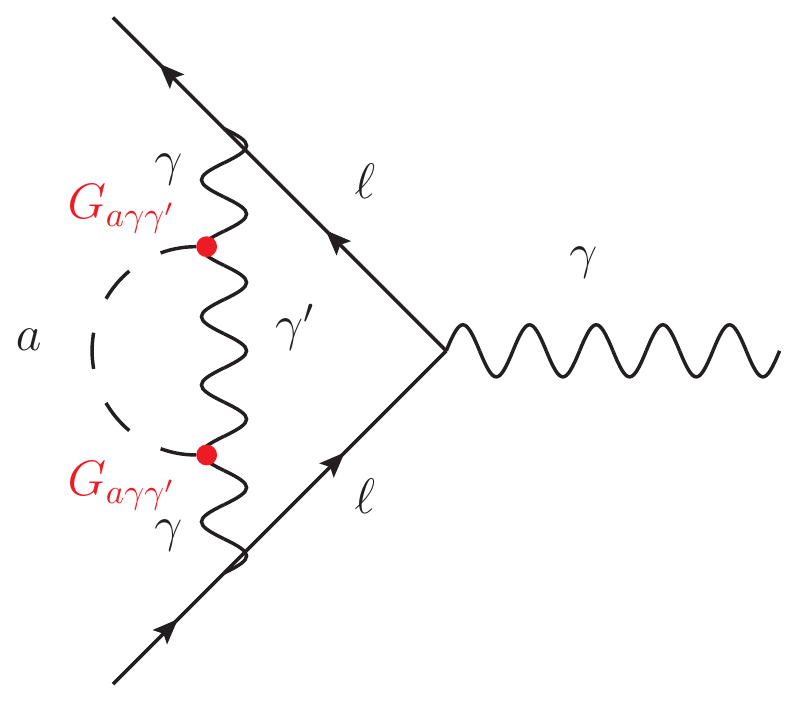}}
 \caption{Two-loop diagram providing the leading contribution to lepton anomalous magnetic moment including $a$-$\gamma$-$\gamma^\prime$ vertices.}
 \label{fig:g_2}
\end{figure}

The dark axion model introduces the new two-loop contribution to the lepton anomalous magnetic moment shown in Fig. \ref{fig:g_2}. The change to lepton $a_\ell = (g-2)/2$ is given by \cite{Jegerlehner:2009ry,Lindner:2016bgg}:
\begin{equation}
 \label{eq:g_minus_2}
 \Delta a_\ell = \frac{\alpha}{\pi} \int_0^1 dx(1-x) \Pi_R(s_x)-\frac{\alpha}{3\pi} c\, m_\ell^2 G_{a\gamma \gamma^\prime}^2,
\end{equation}
where $c$ is a positive free parameter introduced during the renormalization of the $a$-$\gamma$-$\gamma^\prime$ vertex (see App. \ref{app:g_2} for further details),
\begin{equation}
 s_x \equiv -\frac{x^2}{1-x}m_\ell^2,
\end{equation}
and
\bea
\Pi_R(q^2) &=& \Pi(q^2) - \Pi(0) - q^2\Pi^\prime(0)\\
	   &=& \int_0^1 dx \left[ \log \left( \frac{xm_a^2+(1-x)m_{\gamma^\prime}^2}{xm_a^2+(1-x)m_{\gamma^\prime}^2-x(1-x)q^2} \right) \right. \nn \\
	   &&\times(xm_a^2+(1-x)m_{\gamma^\prime}^2-x(1-x)q^2) - \frac{q^2}{6} \Biggr].
\label{eq:PiR}
\eea

While the free parameter $c$ makes the theory unpredictive, both terms that contribute to $\Delta a$ are always negative, and conservative limits can be placed on $G_{a\gamma\gamma^\prime}$ by assuming $c=0$, as non-zero values of $c$ will only magnify the effect of the dark axion portal contribution and correspondingly improve the limits. The current best measurements of the anomalous magnetic moment of the muon come from a muon storage ring at Brookhaven National Laboratory \cite{Mohr:2012tt,Bennett:2002jb,Bennett:2004pv,Bennett:2006fi}. Their measurement exceeds the theoretically predicted value by $3.5\sigma$ \cite{Patrignani:2016xqp},
\begin{equation}
\Delta a_\mu = a_\mu(\text{exp}) - a_\mu(\text{SM}) = (26.8 \pm 7.6)\times10^{-10}.
\end{equation}
The dark axion portal unfortunately exacerbates this disagreement. We place a limit where the SM+dark axion portal increases $\Delta a_\mu$ by $15.2 \times 10^{-10}$, a $5.5\sigma$ disagreement. In the future, the E989 collaboration at FNAL intends to improve on the precision of the current experimental measurement by a factor of three \cite{Grange:2015fou}.

The electron anomalous magnetic moment has been determined most accurately through one-electron quantum cyclotron experiments and measurements of the ratio between the Planck constant and the mass of Rubidium-87 \cite{Bouchendira:2010es,Hanneke:2008tm,Hanneke:2010au}. The theory prediction \cite{Aoyama:2012wj} exceeds experimental measurements by approximately $1\sigma$ \cite{Endo:2012hp},
\begin{equation}
\Delta a_e =  a_e(\text{exp}) - a_e(\text{SM}) = -(1.06\pm0.82)\times10^{-12}.
\end{equation}
The dark axion portal can reduce the disagreement between theory and experiment of the electron anomalous magnetic moment. We place a limit where the dark axion portal contribution overcorrects the difference between the SM and experiment, and $a_e$ disagrees with the experimentally measured value by more than $2\sigma$. Both this contour and that derived for muon $g-2$ are shown in Fig. \ref{fig:limits_2}.

While the two-loop contribution from the dark axion portal cannot resolve the muon $g-2$ discrepancy (because of the wrong sign), the situation could become nontrivial if we allow for the model-dependent contributions from $a$-fermion-fermion Yukawa couplings or $a$-$\gamma$-$\gamma$ coupling ($G_{a\gamma\gamma}$).
As studied in Ref. \cite{Marciano:2016yhf}, the combined effect of Bar-Zee one-loop diagrams and light-by-light and vacuum polarization two-loop diagrams might resolve the muon $g-2$ discrepancy if sufficiently large coupling strengths are allowed.
We will study this more general situation in subsequent works.

\section{Fixed Target Neutrino Experiments}
\label{sec:ftnf}

Fixed target neutrino experiments (FTNEs) impact high intensity proton beams onto thick targets to produce charged mesons, primarily pions and kaons, the decays of which produce neutrinos. FTNEs deliver in excess of $10^{20}$ protons on target (POT) over the life of their running time. While the objective of FTNEs is to study neutrino oscillations, their high intensity and low Standard Model backgrounds are also well suited to searching for hidden sector states with sub-GeV masses \cite{Pospelov:2007mp,Batell:2009di}. This section will consider the sensitivity of LSND and MiniBooNE to the dark axion portal by repurposing published analyses of neutral current elastic scattering. All cross sections in this section were calculated with the assistance of FeynCalc \cite{Shtabovenko:2016sxi,Mertig:1990an}.

\begin{figure}[t]
 \centerline{\includegraphics[width=0.3\textwidth]{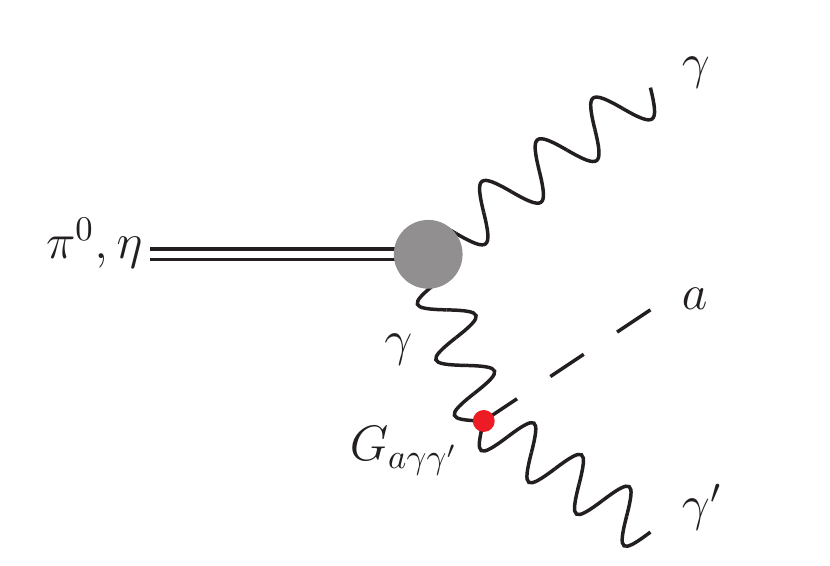}}
 \caption{Decay of the pseudoscalar mesons $\pi^0$ and $\eta$ to $\gamma a \gamma^\prime$ used in the analysis of LSND, MiniBooNE and CHARM. The off-shell internal photon and three body final state suppress the branching ratio.}
 \label{fig:meson_decay}
\end{figure}

Alongside the charged mesons, FTNEs also produce the neutral pseudoscalar mesons $\pi^0$ and $\eta$. The $\pi^0$ is produced in quantities similar to those of the $\pi^+$ and $\pi^-$, while the $\eta$ is produced at a rate suppressed by a factor of 20 to 30 \cite{Jaeger:1974pk,Amaldi:1979zk}. The $a$ and the $\gamma^\prime$ could be produced in radiative decays of the pseudoscalar mesons through the diagram shown in Fig. \ref{fig:meson_decay}. The partial decay width of the decay $\pi^0 \to a \gamma \gamma^\prime$ is given by
\begin{equation}
 \frac{d^2\Gamma}{dm_{12}^2 dm_{23}^2} = \frac{1}{(2\pi)^3}\frac{1}{32m_{\pi^0}^3}\overline{|\mathcal{M}|^2},
\end{equation}
where $m_{ij}^2=(p_i+p_j)^2$ for $i,j=1,2,3$, where particle 1 corresponds to the $\gamma$, particle 2 is the $a$ and particle 3 is the $\gamma^\prime$, and the amplitude is
\begin{align*}
 \overline{|\mathcal{M}|^2} =& \frac{e^4 G_{a\gamma\gamma^\prime}^2}{64 \pi^4 f_\pi^2 m_{23}^4} \bigg[m_{23}^4 \left(m_{12}^2+m_{23}^2-m_a^2-m_{\pi^0}^2 \right)^2 \nonumber\\
&-m_{23}^2\left(m_{23}^2-m_{\pi^0}^2\right) \left(m_{23}^2-m_a^2+m_{\gamma^\prime}^2\right)\nonumber\\ &\times\left(m_{12}^2+m_{23}^2-m_a^2-m_{\pi^0}^2\right)\nonumber\\
 &+\frac{1}{2}\left(m_{23}^2-m_{\pi^0}^2\right)^2\left(m_{23}^2-m_a^2+m_{\gamma^\prime}^2\right)^2\nonumber\\&-m_{23}^2 m_{\gamma^\prime}^2\left(m_{23}^2-m_{\pi^0}^2\right)^2\bigg],
\end{align*}
where $m_e$ is the electron mass, $e = \sqrt{4\pi\alpha_\mathrm{em}}$ and $\alpha_mathrm{em}$ is the fine structure constant. The same expression holds for $\eta$, but with $m_{\pi^0}\to m_{\eta}$ The decay width is suppressed by the kinematics of the three-body final state and the off-shell internal photon propagator. Interestingly, the $\eta$ is far more likely to decay to the dark sector than the $\pi^0$ due to the dependence of the width on the meson mass.

\begin{figure}[t]
 \centerline{\includegraphics[width=0.25\textwidth]{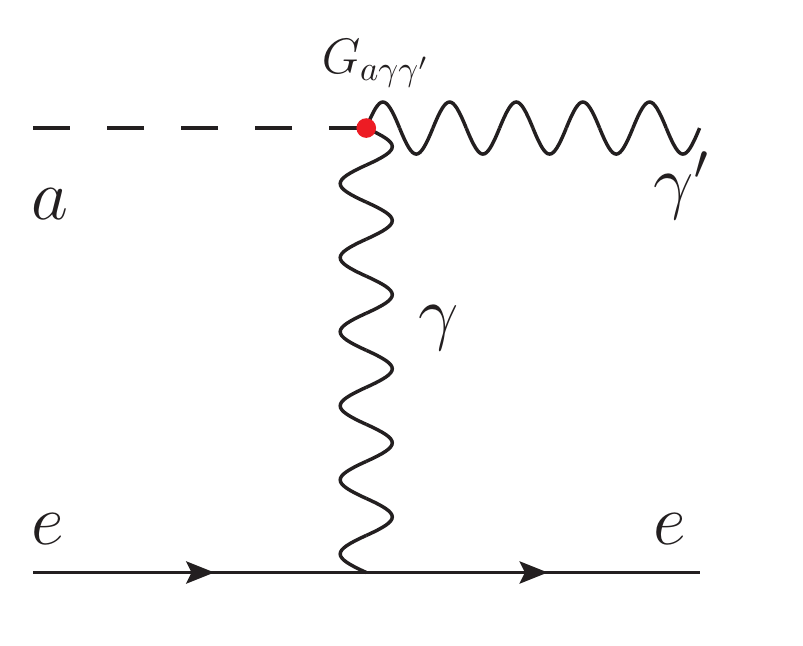}\includegraphics[width=0.25\textwidth]{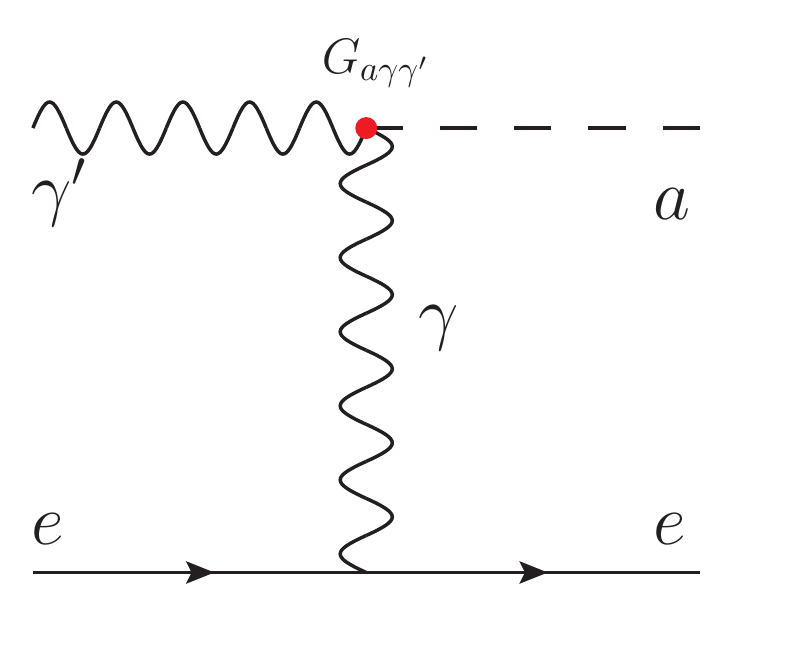}}
 \caption{Inelastic scattering channels observable by MiniBooNE and LSND. The right-hand diagram is always kinematically accessible, while the left-hand diagram requires an energetic $a$ to enable scattering into the higher mass $\gamma^\prime$ state.}
 \label{fig:scatter}
\end{figure}

Both the $a$ and the $\gamma^\prime$ are able to propagate to the neutrino detector where they could be observed through the inelastic scattering channels $ae\to \gamma^\prime e$ and $\gamma^\prime e \to a e$ shown in Fig. \ref{fig:scatter}. The scattering cross section is given by
\begin{equation}
 \frac{d\sigma}{dt} = \frac{1}{64\pi s} \frac{S\Sigma|\mathcal{M}|^2}{|\mathbf{p}_\mathrm{1cm}|^2},
\end{equation}
where $s$ and $t$ are the Mandelstam variables, $\mathbf{p}_\mathrm{1cm}$ is the center of mass momentum of the incoming $a$ or $\gamma^\prime$, $S=\frac{1}{2s+1}$ is a spin symmetry factor, and is equal to $1$ for the $a$ and $1/3$ for the $\gamma^\prime$, and the squared amplitude is given in the limit of $m_a\to0$ by
\begin{align}
 \Sigma{|\mathcal{M}|^2} = -\frac{G_{a\gamma\gamma^\prime}^2 e^2}{6t^2}&\bigg[m_{\gamma^\prime}^4\left(2m_e^2+t\right) -2m_{\gamma^\prime}^2 t\left(m_e^2+s+t\right) \nonumber \\ &+t\left(2[m_e^2-s]^2+2st+t^2\right)\bigg].
\end{align}

\begin{figure}[t]
\subfigure[]{ 
\centerline{\includegraphics[width=0.45\textwidth]{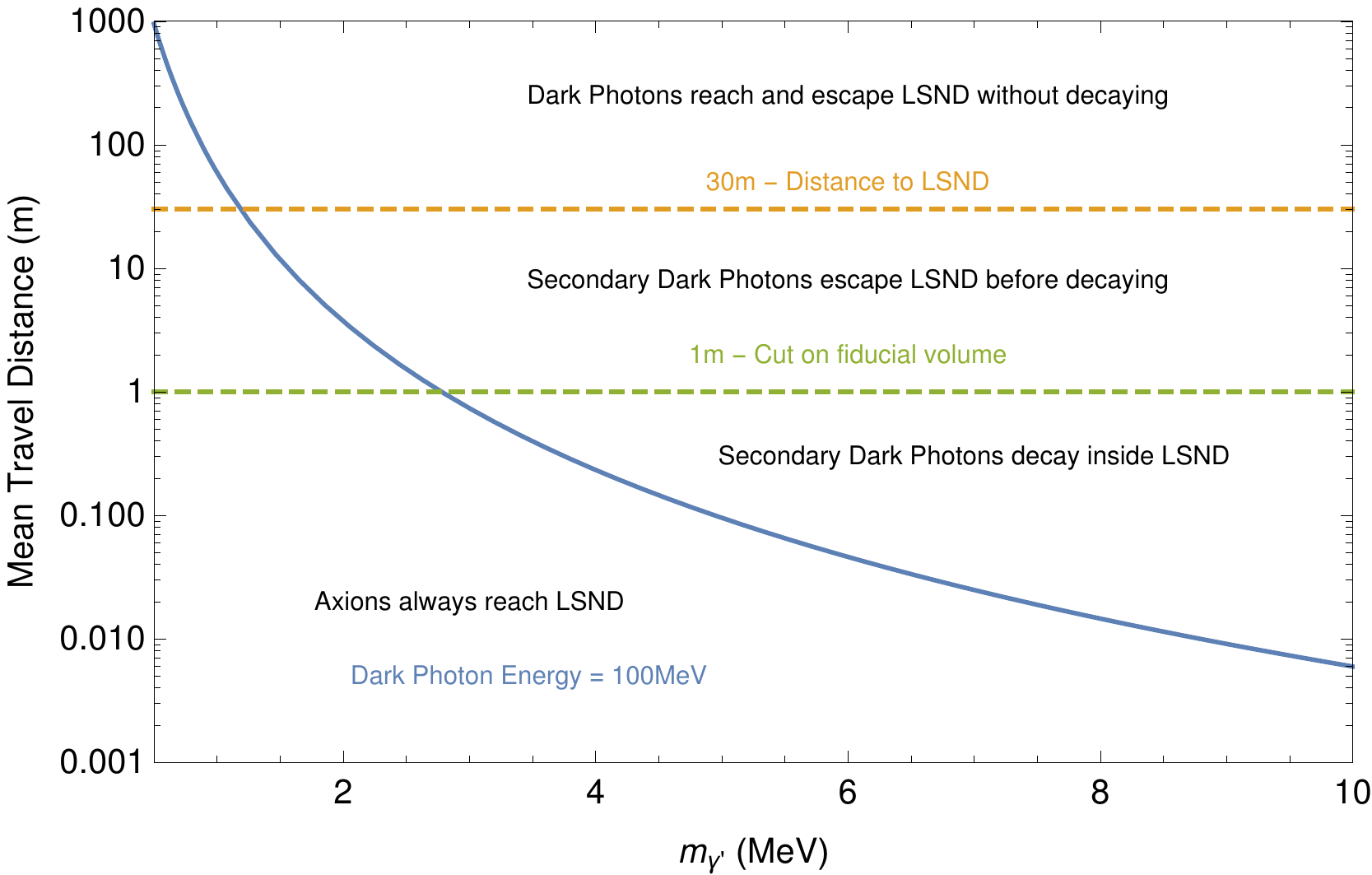}}
}
\subfigure[]{
\centerline{\includegraphics[width=0.45\textwidth]{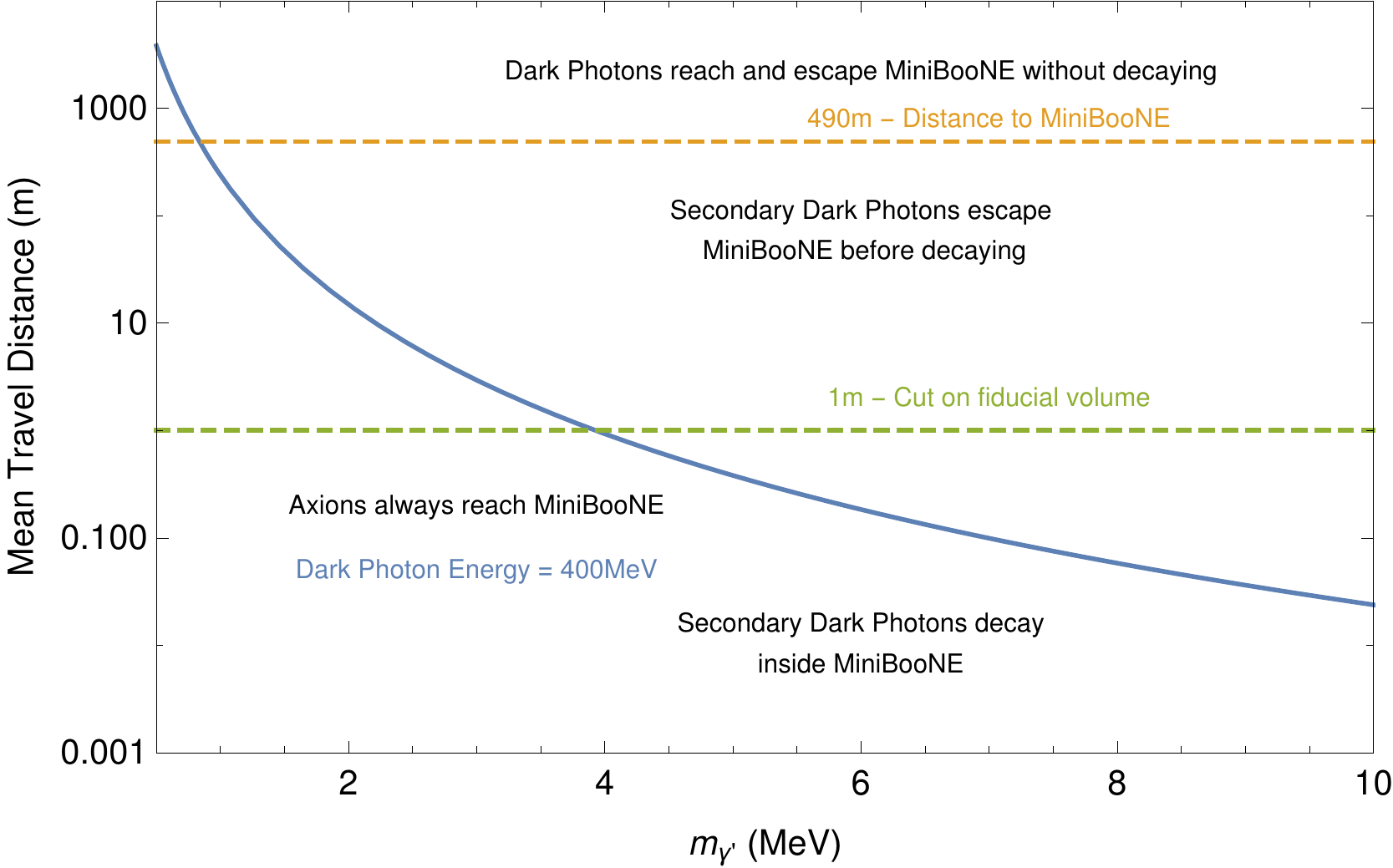}}
}
 \caption{Mean travel distance before decay for characteristic energies of $\gamma^\prime$ at (a) LSND and (b) MiniBooNE. Also marked are relevant distance scales: the distance to the LSND and the MiniBooNE detectors as well as the 1\,m fiducial volume cut. The secondary $\gamma^\prime$ produced in the LSND and MiniBooNE detectors through $ae\to \gamma^\prime e$ scattering have lower energies than those produced in the target.}
 \label{fig:decay_distance}
\end{figure}

Should the $\gamma^\prime$ be sufficiently massive, it will decay to $a\gamma$ before reaching the detector, and only a beam of the long-lived $a$'s will reach the detector. The mass reach of FTNEs is restricted by the kinematics of the inelastic scattering, as the $a$ must be increasingly energetic to scatter into a higher mass state. There is an additional complication in $a$-electron scattering should the $\gamma^\prime$ not escape the detector before its decay. See Fig~\ref{fig:decay_distance} for some examples of the characteristic travel distances before decay. As we will be comparing with Neutral Current Elastic-like analyses to impose limits on the dark axion portal, we will exclude events in which the scattering produces an additional photon.

A modified version of the \textsc{BdNMC} code \cite{deNiverville:2016rqh} was used to simulate both the production and detection of the dark matter signal expected at LSND and MiniBooNE. In the case of LSND, the code takes a momentum distribution of initial mesons chosen based on the experiment and beam energies and produces meson four-momenta by sampling the distribution using an acceptance-rejection algorithm. In the case of MiniBooNE, the code draws four-momenta from a prepared list of sample mesons generated by the MiniBooNE Collaboration. Decays into dark sector particles are generated from the selected meson four-momenta, and the propagation trajectories of those particles are checked for intersection with the detector geometry. If the $\gamma^\prime$ decays before reaching the detector, the resulting decay axion is also checked for intersection. The likelihood of this process is highly dependent on the precise value of $G_{a\gamma\gamma^\prime}$, but adjusting the coupling has only a minor effect on the event rate beyond the expected $G_{a\gamma\gamma^\prime}^4$ scaling for the parameter space of interest, as $\gamma^\prime$ particles are replaced with $a$ particles.

Once a dark sector particle $i$ reaches the detector, it scatters with a probability of 
\begin{equation}
\label{ref:pscatter}
 P_i = \frac{\sigma(E_i)\times L_i}{P_\mathrm{max}},
\end{equation}
where $E_i$ is the energy of the incident particle, $L_i$ is the length of the particle intersection with the detector, $\sigma$ is the scattering cross section and $P_\mathrm{max}$ is the maximum recorded scattering probability. Once a scattering occurs, the differential scattering cross section is sampled with acceptance-rejection, and a set of end state particles is generated and recorded. The total event rate is calculated as
\begin{equation}
\label{eq:ftnf_events}
    N_\mathrm{events} =\frac{N_\mathrm{scatter}}{N_\mathrm{trials}}P_\mathrm{max}\sum_{\alpha=\pi^0,\eta}N_{\alpha}\times \mathrm{Br}(\alpha\to a\gamma\gamma^\prime),
\end{equation}
where $N_\mathrm{scatter}$ is the number of scattering events generated, $N_\mathrm{trials}$ is the total number of attempts that were required to generate those scattering events, and $N_\alpha$ is the total number of mesons of type $\alpha$ produced by the experiment.

Some additional processing was performed while calculating the constraint line in order to determine the probability that the secondary dark photons produced in $ae\to\gamma^\prime e$ scattering escaped the detector undetected. Each experiment excludes a region of the parameter space if it would observe in excess of some number of dark sector events $N_\mathrm{cut}$, where this is largely determined the size of the neutrino signal. The value of $G_{a\gamma\gamma^\prime}$ was calculated iteratively with
\begin{equation}
\label{eq:dark_photon_efficiency}
 G_{a\gamma\gamma^\prime}^{i+1} = \left(\frac{N_\mathrm{cut}}{N_\mathrm{events} \epsilon_i} \right)^{0.25}\times G_{a\gamma\gamma^\prime}^i, 
\end{equation}
where $G_{a\gamma\gamma^\prime}^0$ was the value of the coupling used for the simulation and $\epsilon_i$ is an efficiency factor representing the percentage of dark photons that escaped the detector and $\epsilon_0=1$. The efficiency factor was calculated for each $G_{a\gamma\gamma^\prime}^{i}$ for $i\neq0$ as follows: The probability of a dark photon with energy $E_j$ traveling a set distance $L$ before decaying was averaged over a range of $L$ between 1\,m and the size of the detector itself. The lower bound of 1\,m was chosen to ensure that the dark photon did not decay in any veto regions surrounding the fiducial volume of the detector. The value $\epsilon_i$ was set to the fraction of $\gamma^\prime$ which escaped the detector undetected for a coupling of $G_{a\gamma\gamma^\prime}^i$. This limit could be improved by a more refined treatment of the cut-off, but this would require an analysis using the experiment's own detector Monte Carlo. As implemented, this process provides a conservative estimate of the cutoff, as any refinement would lead to an improvement of the efficiency factor $\epsilon$. A more complicated treatment is also possible by more carefully considering the detector geometry but it would only lead to small changes in the constraint contour as the behavior of the efficiency factor is primarily determined by the cut-off of 1\,m. This iterative process was terminated when $\epsilon_i \left(G_{a\gamma\gamma^\prime}^i\right)^4 N_\mathrm{events}$ differed from $N_\mathrm{cut}$ by less than some tolerance fraction, which we took to be 1\%. Note that the efficiency only becomes important for masses sufficiently large that almost all $\gamma^\prime$ decay before reaching the detector, as otherwise $\epsilon\approx1$. See Fig. \ref{fig:decay_distance} for a visual representation of this effect.

\subsection{LSND}
\label{ssec:lsnd}

LSND was an experiment that ran at Los Alamos Neutron Science Centre from 1994 to 1998 \cite{Athanassopoulos:1996ds, Aguilar:2001ty}. The experiment delivered a total of $1.8\times10^{23}$ POT with a kinetic energy of 798 MeV. The experiment used a 167 tonne mineral oil detector with a diameter of 5.7 meters and a length of 8.3 meters located 30 meters downstream and 4.6 meters below the target\footnote{When calculating the event rate, it is important to note that substantial portions of the detector were excluded from the fiducial volume to serve as cosmic vetoes and improve reconstruction efficiency.}. 

This analysis will be following the lead of previous efforts in Refs. \cite{deNiverville:2011it,Kahn:2014sra} and focus on $a\gamma\gamma^\prime$ production through radiative $\pi^0$ decays, as the $\eta$ is unlikely to be produced in significant numbers. We follow previous work and estimate the $\pi^0$ production rate to be  $N_{\pi^0}=0.06\times$POT$=1.08\times10^{22}$. The Burman-Smith distribution \cite{Burman:1989ds} was used to generate the $\pi^0$ momentum distribution.

For signal, we compare the expected signal from the dark axion portal with the analysis presented in Ref.~\cite{Auerbach:2001wg}, and look for electron recoil events with energies in the range $[15,53]\,\mathrm{MeV}$. We assume a detection efficiency of 16\%, and place a limit on 110 dark axion portal events. Events in which a photon is subsequently produced inside of the detector by the decay of the $\gamma^\prime$ are discarded. The drop in the event rate is reflected by the sudden cutoff in the constraint curve in Fig. \ref{fig:limits_2}, as the distance the dark photon travels before decaying is much smaller than the size of the detector for $m_{\gamma^\prime}\ge2.5\,\mathrm{MeV}$. If some means of ignoring or utilizing the photon produced in the decay of the $\gamma^\prime$ was available, we would expect LSND to be able to place limits on the scenario for $m_{\gamma^\prime}<30\,\mathrm{MeV}$.

\subsection{MiniBooNE}
\label{ssec:miniboone}

MiniBooNE is a fixed target neutrino experiment at Fermi National Accelerator Laboratory (FNAL) that was conducted, in part, to verify the results of the LSND experiment. It ran in on-target mode from 2002 to 2012 with a 70\,cm beryllium target \cite{AguilarArevalo:2008yp,Aguilar-Arevalo:2012fmn,Aguilar-Arevalo:2013nkf}. However, more useful for this analysis was a later 2013-2014 run in which the MiniBooNE experiment ran in off-target mode, directing their proton beam around the target and into the steel beam dump at the end of a 50\,m long open air decay pipe \cite{Aguilar-Arevalo:2017mqx,Aguilar-Arevalo:2018wea}. This dramatically reduced the background from the neutrino signal itself. During this run, the MiniBooNE experiment received $1.86\times10^{20}$ POT with a total energy of 8.9\,GeV. The MiniBooNE detector is a 12\,m diameter sphere filled with 818 tonnes of mineral oil and located 490 meters downstream from the steel beam dump \cite{AguilarArevalo:2008qa}.

A similar analysis to that of LSND can be performed for MiniBooNE, and this work will mirror previous dark matter searches at proton fixed targets \cite{deNiverville:2012ij,deNiverville:2016rqh}. Only $\pi^0$ and $\eta$ decays are considered in this work. While proton bremsstrahlung could contribute, meson decays dominate the hidden sector signal at these experiments for $m_{\gamma^\prime}$ masses below 100\,MeV. The $\pi^0$ and $\eta$ production rates and distributions are drawn from the public data release in the recent MiniBooNE analysis \cite{Aguilar-Arevalo:2018wea}. The total number of $\pi^0$'s produced is calculated to be $N_{\pi^0}=2.5 \times 1.86\times10^{20} = 4.65\times10^{20}$, while the number of $\eta$'s is estimated to be $N_{\pi^0}/30$. The momentum distributions were generated by drawing from the sample $\pi^0$ and $\eta$ positions and momenta supplied in the MiniBooNE data release.

The handling of the signal is similar to the treatment given at LSND, but instead of a cut on electron recoil energy we employ a cut on the electron recoil angle relative to the beamline direction of $\cos\theta_e > 0.99$. This cut removes nearly all of the neutrino background, and the exclusion curve in Fig. \ref{fig:limits_2} was made for 2.3 events with a detection efficiency of 35\%. This exclusion curve demonstrates the same sharp cutoff as LSND, but appears at a larger mass due to the higher energy, effectively extending the lifetime of the $\gamma^\prime$ due to the larger boost factor.

\begin{figure}[t]
 \centerline{\includegraphics[height=0.33\textwidth]{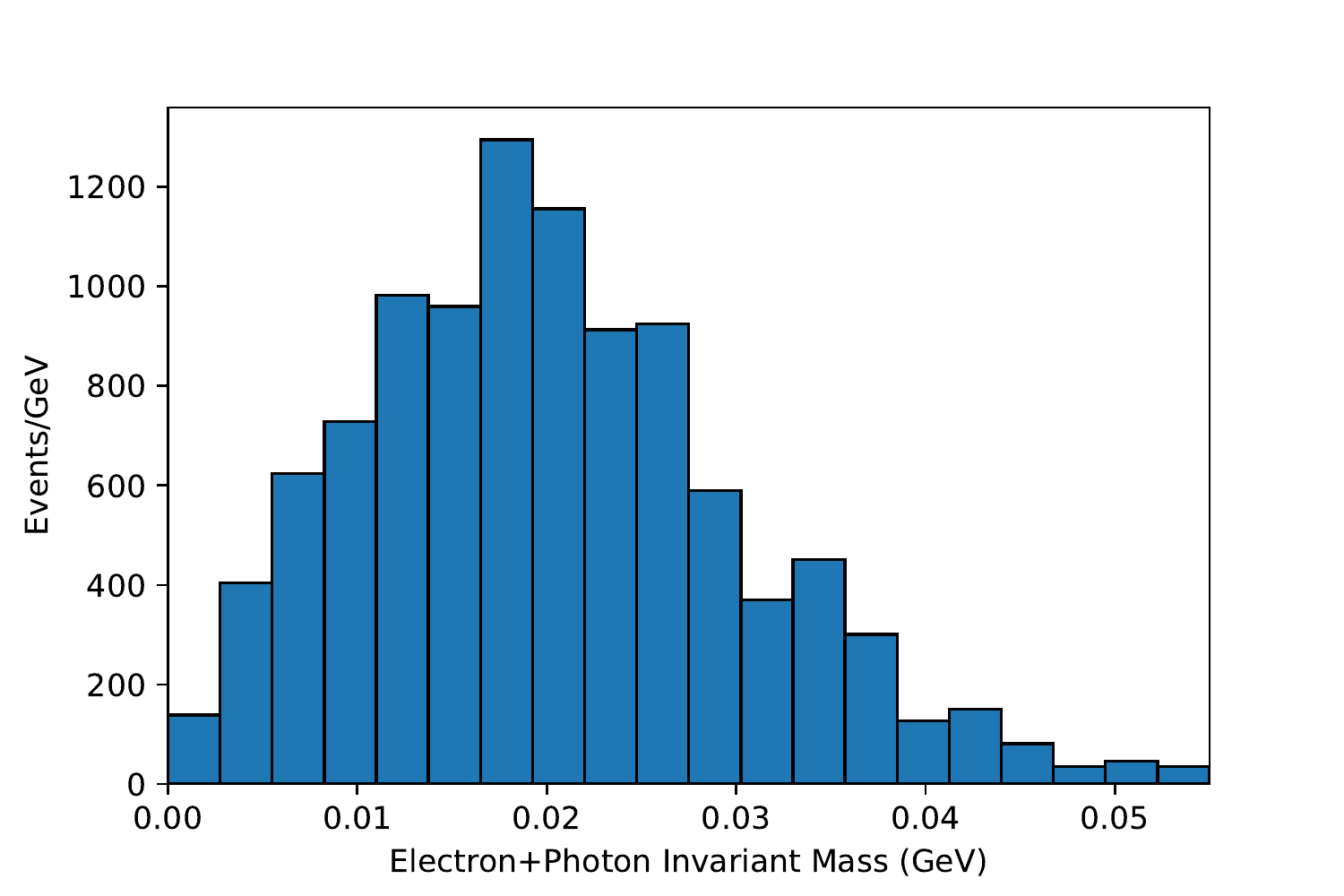}}
 \caption{Invariant mass spectrum of the photon and recoil electron produced through $ae\to \gamma^\prime e \to \gamma a e$, where $s = (p_\gamma+p_e)^2$. The histogram shown is for the MiniBooNE off-target run, in which the proton beam was impacted on the 50\,m steel beam dump, with $m_{\gamma^\prime}=3\,\mathrm{MeV}$ and $G_{a\gamma\gamma^\prime}=0.01$. This invariant mass range is too small to survive the cuts placed by the MiniBooNE $\pi^0$ reconstruction, as it requires $\sqrt{s}=[80,200]\,\mathrm{MeV}$. A search of the $m_{\gamma\gamma}$ spectrum for smaller invariant masses would be sensitive to this scenario.}
 \label{fig:inv_mass_spec}
\end{figure}

An interesting quirk of the MiniBooNE detector is its difficulty in differentiating electrons and photons. This leads to the possibility of extending the analysis to higher masses by reconstructing the recoil electron and secondary photon produced through $ae\to \gamma^\prime e \to \gamma a e$ as a photon pair produced through $\pi^0\to\gamma\gamma$. The relevant MiniBooNE analysis \cite{Aguilar-Arevalo:2018wea} requires $\sqrt{s} \in [80,200]\,\mathrm{MeV}$ and the invariant mass of the recoil electron and decay photon produced through $a$-electron scattering is too small to survive the cuts, as shown for a sample in Fig. \ref{fig:inv_mass_spec}. Were further analysis focused on $\sqrt{s} \le 80\,\mathrm{MeV}$ performed, it is possible that the constraints could be extended to larger masses. 

\section{Beam Dump Experiments}
\label{sec:beamdump}

\subsection{CHARM}
\label{ssec:charm}

Beam dump experiments impact high intensity beams of protons and electrons on thick targets in order to generate weakly coupled particles, either directly or through the subsequent decays of heavy particles in a downstream decay volume. These particles then travel tens to hundreds of meters through beam stops, dirt and air before being detected through either their decay into Standard Model particles or, in the case of neutrinos, their scattering interactions with the detector material. This section considers the sensitivity to the dark axion portal of two CHARM searches for heavy neutrinos.

Both analyses follow Refs. \cite{Gninenko:2011uv,Gninenko:2012eq} and consider production through $\pi^0$ and $\eta$ decays. Following the previous works, the ratio of $\eta$ mesons to $\pi^0$ mesons is taken to be $N_{\pi^0}/N_\eta=0.078$. While $\eta^\prime$ decays could also be considered in order reach larger $m_{\gamma^\prime}$, it is the rapid decline in the lifetime of the $\gamma^\prime$ with increasing mass and the distance to the detector that determines the $m_{\gamma^\prime}$ reach of the CHARM experiment rather than the phase space, and branching ratio, available in meson decays. As it is produced in smaller quantities than the $\eta$ and $\pi^0$, we will neglect the $\eta^\prime$ contribution in this work. The overall $\pi^0$ production rate as well as its momentum distribution was calculated with the BMPT distribution for a 300\,cm copper target \cite{Bonesini:2001iz}. The BMPT distribution is also used for the $\eta$ momentum distribution, as the $\pi^0$ and $\eta$ momentum distributions are expected to be quite similar. 

The CHARM fine-grain target calorimeter is composed of 72 marble plates with a thickness of 8\,cm, spaced 20\,cm apart with scintillation counters and proportional drift tubes inserted in the intervening space \cite{Dorenbosch:1987pn}. The center of the detector is located 487.3\,m downstream from the target. The fiducial volume is made up of a cross-section of 2.4 $\times$ 2.4\,$\mathrm{m}^2$, beginning at target plane 3 and ending at target plane 59, with the first three planes serving as vetoes for particle production in the upstream muon spectrometer. The last 14 planes are used for shower measurements, as particles produced in these planes might escape the detector before depositing sufficient energy to correctly reconstruct the shower.

We now consider two analyses that we will label (1) and (2). In analysis (1), the detector was used in a heavy neutrino decay search \cite{Bergsma:1983rt} with 7.1$\times10^{18}$ POT. The analysis searched for electron-positron pairs produced through neutrino decays, where the resulting electromagnetic showers possessed $E\in[7.5,50]\,\mathrm{GeV}$ and $E^2\theta^2<0.54\,\mathrm{GeV}^2$. Of particular interest for the dark axion portal, single photon emission such as that produced through the decay $\gamma^\prime\to a\gamma$, could also survive these cuts. This is the dominant decay channel for the $\gamma^\prime$ in the dark axion portal. The analysis attributed $1\pm 49$ events to heavy neutrino decays, and we conservatively exclude regions of the parameter space that generate more than 99 events.

\begin{figure}[t]
 \centerline{\includegraphics[height=0.3\textwidth]{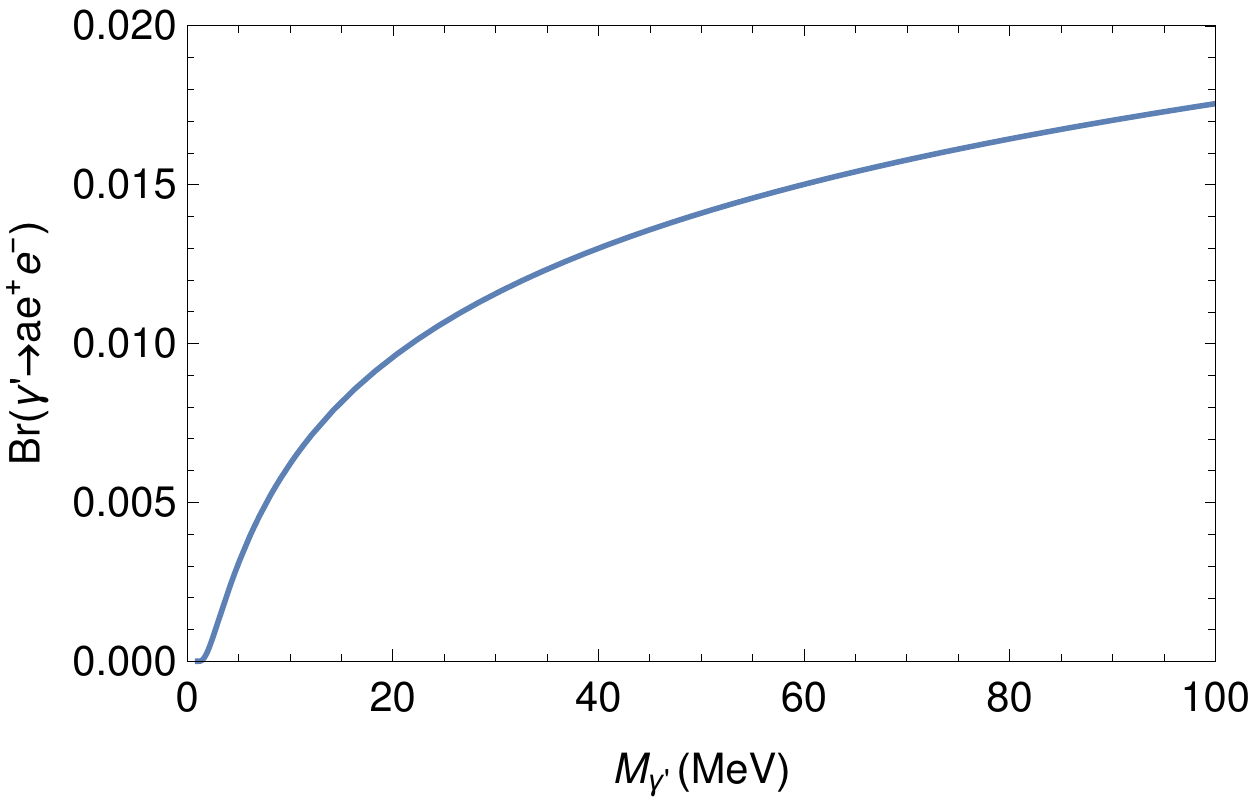}}
 \caption{The branching ratio of $\gamma^\prime \to a e^+ e^-$, only considering decays through the $a\gamma\gamma^\prime$ vertex.}
 \label{fig:A_to_aee}
\end{figure}

In analysis (2), a search for electron positron pair production was conducted in a 35\,m long decay volume with a cross section of 3 $\times$ 3\,$\mathrm{m}^2$ \cite{Bergsma:1985is}. This decay region is parallel to the neutrino beamline, but offset by 5\,m. This data was used to constrain dark photon decays in Ref. \cite{Gninenko:2012eq}.  This search required the electron-positron pair to possess greater than $3\,\mathrm{GeV}$ of energy. The decay $\gamma^\prime\to a e^+ e^-$ is rare compared to the radiative decay channel, with a branching fraction rising to nearly 2\% for $m_{\gamma^\prime} \sim 100\,\mathrm{MeV}$ (see Fig.~\ref{fig:A_to_aee}). This analysis was only performed on 2.4$\times10^{18}$ POT, and recorded zero background events. We exclude regions of the parameter space for which this analysis would have observed more than 2.3 events.

Both analyses described above were simulated with a modified version of the \textsc{BdNMC} code, previously described in Sec.~\ref{sec:ftnf}. The event rate calculation is similar to that shown in Eq.~\eqref{eq:ftnf_events}, but with $N_\mathrm{scatters} \to N_\mathrm{decays}$ and a different calculation of the event probability for some $\gamma^\prime$ $i$:
\begin{equation}
\label{eq:dec_prob  }
 P_\mathrm{decay,i} = \mathrm{Br}(\gamma^\prime\to X)\left[\exp\left(-\frac{L_{1,i} E_i}{c\tau m}\right) - \exp\left(-\frac{L_{2,i} E_i}{c\tau m}\right)\right],
\end{equation}
where $\mathrm{Br}(\gamma^\prime\to X)$ is the probability of the $\gamma^\prime$ decaying via the channel of interest, with $X=a\gamma$ for analysis (1) and $X=ae^+e^-$ for analysis (2), $E_i$ is the $\gamma^\prime$ energy and $L_{1,i},L_{2,i}$ are the distances from the center of the production target at which $\gamma^\prime$ $i$'s trajectory enters and exits the target, respectively.

The constraints placed by CHARM on the dark axion portal are shown in Fig.~\ref{fig:limits_2}. Due to a combination of the far larger radiative branching ratio and larger exposure, analysis (1) places stronger limits than analysis (2) and the contour shown is entirely due to the $\gamma^\prime\to a \gamma$ signal. Note that unlike searches for $e^+ e^-$ pairs, the contour shown extends to masses below $m_{\gamma^\prime} = 2 m_e$.

\subsection{Electron Beam Dumps}
\label{ssec:ebd}

Also considered were electron beam dumps, with the E137 experiment \cite{Bjorken:1988as} taken as a test case. E137 was a beam dump experiment that searched for metastable hidden sector particles, generating them through bremsstrahlung by impacting 30 C of 20 GeV electrons on an Aluminum target. The particles then travelled 383\,m to a detector, the exact makeup of which changed between the two runs of the experiment. The signal required for detection, more than 1\,GeV of energy deposited in the electromagnetic calorimeters \cite{Batell:2014mga}, is easily satisfied by the dark axion portal through inelastic scattering with electrons in the detector. The bremsstrahlung cross section $eN \to eN\gamma^\prime a$ was calculated with CalcHEP 3.6.27 \cite{Pukhov:1999gg,Belyaev:2012qa} and CT10 parton distribution functions \cite{Lai:2010vv}. Unfortunately, the cross section appears to be far too small (on the order of several hundred pb) to generate any events without a prohibitively large value of $G_{a\gamma\gamma^\prime}$. This heavy suppression should extend to all electron beam dumps, and without some additional enhancement to the production rate, they are unable to place strong limits on the scenario.

\section{Summary and Discussion}
\label{sec:summary}
We studied implications of the dark axion portal for the lepton $g-2$, $B$-factories, fixed target neutrino experiments and the CHARM experiment in the $m_a \ll m_{\gamma'}$ limit. We focused on the dark photon masses for which $B$-factories are most sensitive, roughly from 1\,MeV to 10\,GeV, and we restricted our plotted results to this window in Fig. \ref{fig:limits_2}. BaBar and Belle-II have too little energy to produce an on-shell $\gamma^\prime$ with a mass much larger than 10.2\,GeV. It is important to consider the lifetime of the $\gamma^\prime$ for $m_{\gamma^\prime}$ below a hundred MeV, as it becomes increasingly likely that the dark photon will escape the detector before decaying in an observable fashion. This is reflected in the gradual decline in the sensitivity for smaller masses as the $\gamma^\prime$ lifetime becomes comparable to the size of the BaBar and Belle-II experiments. The mass reach could be extended in both directions by considering off-shell dark photon production through the process $e^+ e^- \to a \gamma^{\prime*} \to a a \gamma$, though this cross section is suppressed by several orders of magnitude relative to $e^+ e^- \to a \gamma^\prime$. 

The finite lifetime of the $\gamma^\prime$ is exploited in the analysis of the CHARM experiment, where we search for observable decays of the $\gamma^\prime$. Both the $\gamma^\prime \to a \gamma$ and $\gamma^\prime \to a e^+ e^-$ were considered, and the former mode was found to provide the best constraints. Note that the constraint from the CHARM experiment possesses both an upper and lower bound, as is characteristic of searches for rare decays in beam dumps. These curves are determined in large part by the lifetime of the $\gamma^\prime$, which scales with $G_{a\gamma\gamma^\prime}^{-2}$. The optimal signal is found when the mean travel distance of the $\gamma^\prime$ is approximately equal to the distance to the detector. If $G_{a\gamma\gamma^\prime}$ is too large, the lifetime is short and it decays before reaching the detector. The upper bound slopes downward because the lifetime also declines with increasing mass. If $G_{a\gamma\gamma^\prime}$ is too small, the $\gamma^\prime$ will be likely to propagate far beyond the detector, reducing its probability of decaying inside the detector. The lower bound is also affected by the overall $\gamma^\prime$ production rate, which declines as $G_{a\gamma\gamma^\prime}^2$. It is the product of these two effects that determines the lower bound of the CHARM exclusion.

The LSND and MiniBooNE analyses are greatly weakened by the short-lived $\gamma^\prime$, as only $a$ particles are sufficiently long-lived to reach the detector for $m_{\gamma^\prime}\ge$~few\,MeV. The $a$ scatters inelastically into $\gamma^\prime$,  the radiative decay of which changes the observed signal. The cutoff in the sensitivity observed in Fig.~\ref{fig:limits_2} appears when the $\gamma^\prime$ is unlikely to escape the detector before decaying. 

We also investigated the effects on both the muon and electron $g-2$ in a conservative manner, though they only imposed meaningful constraints at relatively low masses. These limits become stronger at low masses, as the contribution from the internal $\gamma^\prime-a$ loop is suppressed by a large $m_{\gamma^\prime}$.

We now move on to possible extensions of this work. While we have restricted our attention to monophoton searches at asymmetric $B$-factories, $e^+e^-$ colliders could potentially probe the scenario in other ways. As mentioned in Ref. \cite{Marciano:2016yhf} in the context of axion-like particles, $e^+e^- \to e^+e^-+a \gamma^\prime$ is an intriguing channel, with final states ranging from $e^+e^- + \text{missing energy}$ to $e^+e^- + \text{multiple photons}$ depending on the lifetimes of the dark particles, but has yet to see an experimental analysis.

Evidence of $\gamma \to a\gamma^\prime$ conversion may also be found in radiative meson decays, but the rapid decay of the $\gamma^\prime$ complicates the signal for larger values of $m_{\gamma^\prime}$. For long-lived $\gamma^\prime$'s, we can compare the limit of Br$(\pi^0 \to \gamma \nu \bar \nu)<6\times 10^{-4}$ placed by Ref. \cite{Atiya:1992sm} to the branching fraction of $\pi^0 \to a\gamma\gamma^\prime$, a decay with a similar end-state. Unfortunately, this branching fraction is quite small, and the possible limit of $G_{a\gamma\gamma^\prime}\gsim1\,\gev^{-1}$ is not competitive with those placed by electron or muon $g-2$. For a short-lived $\gamma^\prime$, the signal would be $\pi^0 \to \gamma\gamma +\text{Missing Energy}$, which would require a measurement of the invariant mass distribution of the end-state photons. Limits could also be derived by comparing $K^+ \to \pi^+ \nu \bar \nu$ \cite{Artamonov:2008qb} to $K^+ \to \pi^+ (\gamma^\star \to a\gamma^\prime)$ or $\phi \to \pi^0 \gamma$ to $\phi \to \pi^0 a(\gamma^\prime \to a\gamma)$ \cite{Achasov:2000zd}. For much larger masses, one could look to the Higgs decay $H \to \gamma \gamma^\star \to \gamma\gamma + \text{Missing Energy}$.

Future directions of interest involve exploring the implications of a long-lived $\gamma^\prime$ more thoroughly, as planned beam dump experiments such as SHiP \cite{Anelli:2015pba} could be sensitive to monophotons produced through $\gamma^\prime \to a\gamma$. In the case of very long-lived dark photons, inelastic $a$ or $\gamma^\prime$ scattering inside the detectors of future fixed target neutrino experiments such as those associated    with the Short Baseline Neutrino program \cite{Tufanli:2016hyo}, and reactor neutrino experiments \cite{Ko:2016owz,Park:2017prx} could also be useful search avenues. Missing momentum/energy experiments such as NA64 \cite{Banerjee:2017hhz} provide another probe of the parameter space that may be worth consideration. The constraints from $a$-electron scattering at fixed target experiments should also be extended to consider the effects of the $a$-$\gamma$-$\gamma$ vertex, though in many cases this will resemble a rescaling of existing limits on axion-like particles coupled predominantly to photons.

For masses below a few $\mev$, constraints from stellar cooling and supernovae become interesting, as both the $a$ and the $\gamma^\prime$ provide potential carriers for additional energy loss \cite{Dolan:2017osp,Dreiner:2013mua}, as well as production from the sun \cite{Redondo:2013lna}. It should be noted that the production is suppressed from standard dark photon or axion-like particle searches by the requirement that both an $a$ and a $\gamma^\prime$ are produced simultaneously. A more complete treatment of this limit would require the inclusion of the $a$-$\gamma$-$\gamma$ vertex, as even at a suppressed rate, $a$ reabsorption would have a significant effect on stellar energy loss due to dark particle emission. The $\gamma^\prime$ could escape, but this signal is suppressed by decay to an $a-\gamma$ pair before escape, or inelastical scattering into an $a$ through $\gamma$ mediated interactions with stellar material. 

\begin{acknowledgments}
This work was supported in part by IBS (Project Code IBS-R018-D1), NRF Strategic Research Program (NRF-2017R1E1A1A01072736), NRF Basic Science Research Program (NRF-2016R1A2B4008759) and CAU Research Grants in 2018. We would like to thank Tyler Thornton and Richard Van De Water for helpful discussions regarding the MiniBooNE experiment.
HL appreciates hospitality of the Pitt-PACC at University of Pittsburgh during the Light Dark World International Forum 2017.
\end{acknowledgments}

\appendix

\section{\boldmath Renormalization and Lepton $g-2$}
\label{app:g_2}
\begin{figure}[t]
 \centerline{\includegraphics[width=0.3\textwidth]{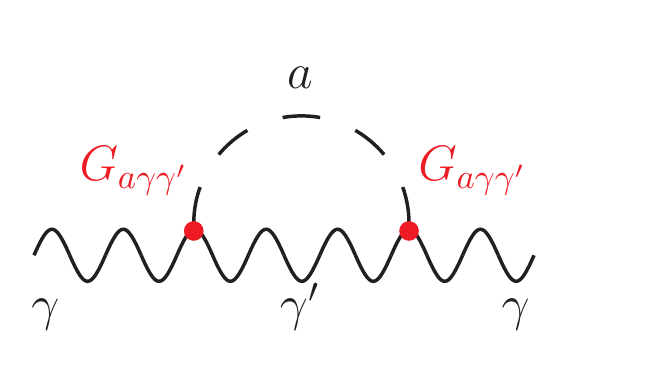}}
 \caption{New photon vacuum polarization diagram that contributes to $\Pi^{\mu\nu}(q^2)$ 
 with the introduction of the $a$-$\gamma$-$\gamma^\prime$ vertex.}
\label{fig:vac_pol}
\end{figure}

As mentioned in Sec. \ref{ssec:g_2}, the dark axion portal introduces a new two-loop contribution (see Fig. \ref{fig:g_2}) to the lepton $g-2$. This contribution is not renormalizable, and we will require additional counter terms to eliminate the new divergences. The first step is to calculate the sub-diagram in Fig. \ref{fig:vac_pol}, the dark axion portal contribution to the photon vacuum polarization,
\begin{equation}
\begin{split}
 i\Pi^{\mu\nu}&= -\frac{1}{F^2} \int \frac{d^4}{(2\pi)^4}\frac{\epsilon^{\sigma \mu \rho \delta} \epsilon_\sigma^{\nu\beta\gamma} k_\rho q_\delta k_\beta q_\gamma}{[k^2-m_{\gamma^\prime}^2][(k+q)^2-m_a^2]}\\
    &\equiv i \Pi(q^2)(q^2 g^{\mu\nu} - q^\mu q^\nu)
\end{split}
\end{equation}
where
\begin{equation}
\begin{split}
 &\Pi(q^2) =  \frac{G^2_{a\gamma\gamma^\prime}}{(4\pi)^2} \int_0^1 dx [x m_a^2+(1-x)m_{\gamma^\prime}^2-x(1-x)q^2] \\ &\left(\frac{2}{\epsilon} - \gamma + \log(4\pi) - \log(x m_a^2 + (1-x)m_{\gamma^\prime}^2 - x(1-x)q^2) \right) .
\end{split}
\label{eq:Pi}
\end{equation}

The integral in Eq. (\eqref{eq:Pi}) is quadratically divergent, and we will need to add an additional quadratically divergent term to the lagrangian to cancel the infinities,
\begin{equation}
 \mathcal{L}_\text{dark axion portal} \ni \frac{c}{4\Lambda^2} \partial_\rho F_{\mu \nu} \partial^\rho F^{\mu \nu},
 \label{eq:g_2_lag}
\end{equation}
where $c$ is a free parameter, and $\Lambda$ is some cut-off scale. The most straightforward approach (and the one we will adopt) would be to set $\Lambda = G_{a\gamma \gamma^\prime}^{-1}$ and use $G_{a\gamma \gamma^\prime}^{-1}$ as the cut-off scale, although it is not mandatory.

In this case, the photon propagator should be modified. If we keep all possible interactions and corrections to $\mathcal{O}(G_{a\gamma \gamma^\prime}^2)$, the quadratic lagrangian for the photon is given by
\begin{equation}
    \begin{split}
    \mathcal{L}_\text{kin} = -&\frac{1}{4} F_{\mu \nu}F^{\mu \nu} + \frac{c G_{a \gamma\gamma^\prime}^2}{4} \partial_\rho F_{\mu \nu} \partial^\rho F^{\mu \nu} \\
    -& \frac{1}{2\xi}\left[(\partial_\mu A^\mu)^2 - cG_{a \gamma\gamma^\prime}^2(\partial_\rho \partial_\mu A^\mu)^2\right]
    \end{split}
\end{equation}
in the $\xi-$gauge.
This can be rewritten as
\begin{equation}
 \frac{1}{2} A_\mu\left[ \partial^2 g^{\mu\nu} - (1-\frac{1}{\xi}\partial^\mu \partial^\nu) \right]\left[1+c G_{a \gamma\gamma^\prime}^2\right]A_\nu,
\end{equation}
up to total divergences. The photon propagator is written as
\begin{equation}
 \frac{-i}{(k^2+i\epsilon)[1-c G_{a\gamma\gamma^\prime}^2 k^2]} \left[g_{\mu \nu}+(1-\xi) k \right] . 
\end{equation}
Note that the propagating part of the propagator can be rewritten as
\begin{equation}
\label{eq:prop}
 \frac{1}{k^2[1-c G_{a\gamma\gamma^\prime}k^2]} = \frac{1}{k^2} - \frac{1}{k^2 - \frac{1}{c G_{a\gamma \gamma^\prime}^2}}
\end{equation}
from which we see that the quartic derivative of the photon field plays the role of the Pauli-Villars regulator, which introduces a `ghost' with a mass term $(c G_{a\gamma \gamma^\prime}^2)^{-1}$. In order to prevent the super-luminal propagation of the ghost, we require $c>0$.

It is interesting to note that the non-renormalizable term of Eq. (\eqref{eq:g_2_lag}) has parallels with the Lee-Wick Standard Model . In this model, by putting the Pauli-Villars regulator as Eq. (\eqref{eq:prop}), the degree of divergence in the loop diagram is reduced. As a result, QED becomes UV finite \cite{Lee:1969fy,Lee:1970iw}, and when extended to the SM, the quadratic divergence in the Higgs mass correction is removed \cite{Grinstein:2007mp}. In order to make the theory unitary, the integration contour in the Feynman diagram is modified, at the price of which the causality is violated microscopically \cite{Cutkosky:1969fq,Lee:1971ix,Coleman:1970lec}.

With the inclusion of the additional counter term, we can write the renormalized form of $\Pi(q^2)$,
\begin{equation}
 \Pi_R(q^2) = \Pi(q^2) - \Pi(0) - q^2\Pi^\prime(0)
\end{equation}
where $\Pi_R(q^2)$ is finite, and $\Pi^\prime(0) =\left. \frac{d \Pi(q^2)}{dq^2} \right|_{q^2=0}$.

The expression for $\Pi_R(q^2)$ can be found in Eq. (\eqref{eq:PiR}), as well as its application in calculating electron and muon $g-2$.



\begin{thebibliography}{99}
\bibitem{Aad:2012tfa} 
  G.~Aad {\it et al.} [ATLAS Collaboration],
  Phys.\ Lett.\ B {\bf 716}, 1 (2012)
  doi:10.1016/j.physletb.2012.08.020
  [arXiv:1207.7214 [hep-ex]].

\bibitem{Chatrchyan:2012xdj} 
  S.~Chatrchyan {\it et al.} [CMS Collaboration],
  Phys.\ Lett.\ B {\bf 716}, 30 (2012)
  doi:10.1016/j.physletb.2012.08.021
  [arXiv:1207.7235 [hep-ex]].
  
\bibitem{Patrignani:2016xqp} 
  C.~Patrignani {\it et al.} [Particle Data Group],
  Chin.\ Phys.\ C {\bf 40}, no. 10, 100001 (2016).
  doi:10.1088/1674-1137/40/10/100001
  
\bibitem{Hall:2009bx} 
  L.~J.~Hall, K.~Jedamzik, J.~March-Russell and S.~M.~West,
  JHEP {\bf 1003}, 080 (2010)
  doi:10.1007/JHEP03(2010)080
  [arXiv:0911.1120 [hep-ph]].
  
\bibitem{Essig:2013lka} 
  R.~Essig {\it et al.},
  arXiv:1311.0029 [hep-ph].
  
\bibitem{Alexander:2016aln} 
  J.~Alexander {\it et al.},
  arXiv:1608.08632 [hep-ph].
  
\bibitem{ArkaniHamed:2008qn} 
  N.~Arkani-Hamed, D.~P.~Finkbeiner, T.~R.~Slatyer and N.~Weiner,
  Phys.\ Rev.\ D {\bf 79}, 015014 (2009)
  doi:10.1103/PhysRevD.79.015014
  [arXiv:0810.0713 [hep-ph]].

\bibitem{Nomura:2008ru} 
  Y.~Nomura and J.~Thaler,
  Phys.\ Rev.\ D {\bf 79}, 075008 (2009)
  doi:10.1103/PhysRevD.79.075008
  [arXiv:0810.5397 [hep-ph]].

\bibitem{Holdom:1985ag} 
  B.~Holdom,
  Phys.\ Lett.\  {\bf 166B}, 196 (1986).
  doi:10.1016/0370-2693(86)91377-8
  
\bibitem{Davoudiasl:2012ag} 
  H.~Davoudiasl, H.~S.~Lee and W.~J.~Marciano,
  Phys.\ Rev.\ D {\bf 85}, 115019 (2012)
  doi:10.1103/PhysRevD.85.115019
  [arXiv:1203.2947 [hep-ph]].

\bibitem{Davoudiasl:2012qa} 
  H.~Davoudiasl, H.~S.~Lee and W.~J.~Marciano,
  Phys.\ Rev.\ Lett.\  {\bf 109}, 031802 (2012)
  doi:10.1103/PhysRevLett.109.031802
  [arXiv:1205.2709 [hep-ph]].

\bibitem{Davoudiasl:2013aya} 
  H.~Davoudiasl, H.~S.~Lee, I.~Lewis and W.~J.~Marciano,
  Phys.\ Rev.\ D {\bf 88}, no. 1, 015022 (2013)
  doi:10.1103/PhysRevD.88.015022
  [arXiv:1304.4935 [hep-ph]].

\bibitem{Lee:2013fda} 
  H.~S.~Lee and M.~Sher,
  Phys.\ Rev.\ D {\bf 87}, no. 11, 115009 (2013)
  doi:10.1103/PhysRevD.87.115009
  [arXiv:1303.6653 [hep-ph]].

\bibitem{Davoudiasl:2014kua} 
  H.~Davoudiasl, H.~S.~Lee and W.~J.~Marciano,
  Phys.\ Rev.\ D {\bf 89}, no. 9, 095006 (2014)
  doi:10.1103/PhysRevD.89.095006
  [arXiv:1402.3620 [hep-ph]].

\bibitem{Kong:2014jwa} 
  K.~Kong, H.~S.~Lee and M.~Park,
  Phys.\ Rev.\ D {\bf 89}, no. 7, 074007 (2014)
  doi:10.1103/PhysRevD.89.074007
  [arXiv:1401.5020 [hep-ph]].

\bibitem{Kim:2014ana} 
  D.~Kim, H.~S.~Lee and M.~Park,
  JHEP {\bf 1503}, 134 (2015)
  doi:10.1007/JHEP03(2015)134
  [arXiv:1411.0668 [hep-ph]].

\bibitem{Davoudiasl:2015bua} 
  H.~Davoudiasl, H.~S.~Lee and W.~J.~Marciano,
  Phys.\ Rev.\ D {\bf 92}, no. 5, 055005 (2015)
  doi:10.1103/PhysRevD.92.055005
  [arXiv:1507.00352 [hep-ph]].

\bibitem{Kaneta:2016wvf} 
  K.~Kaneta, H.~S.~Lee and S.~Yun,
  Phys.\ Rev.\ Lett.\  {\bf 118}, no. 10, 101802 (2017)
  doi:10.1103/PhysRevLett.118.101802
  [arXiv:1611.01466 [hep-ph]].

\bibitem{Choi:2016kke} 
  K.~Choi, H.~Kim and T.~Sekiguchi,
  Phys.\ Rev.\ D {\bf 95}, no. 7, 075008 (2017)
  doi:10.1103/PhysRevD.95.075008
  [arXiv:1611.08569 [hep-ph]].

\bibitem{Kaneta:2017wfh} 
  K.~Kaneta, H.~S.~Lee and S.~Yun,
  Phys.\ Rev.\ D {\bf 95}, no. 11, 115032 (2017)
  doi:10.1103/PhysRevD.95.115032
  [arXiv:1704.07542 [hep-ph]].
    
\bibitem{Agrawal:2017eqm} 
  P.~Agrawal, G.~Marques-Tavares and W.~Xue,
  JHEP {\bf 1803}, 049 (2018)
  doi:10.1007/JHEP03(2018)049
  [arXiv:1708.05008 [hep-ph]].

\bibitem{Kitajima:2017peg} 
  N.~Kitajima, T.~Sekiguchi and F.~Takahashi,
  Phys.\ Lett.\ B {\bf 781}, 684 (2018)
  doi:10.1016/j.physletb.2018.04.024
  [arXiv:1711.06590 [hep-ph]].

\bibitem{Choi:2018dqr} 
  K.~Choi, H.~Kim and T.~Sekiguchi,
  arXiv:1802.07269 [hep-ph].

\bibitem{Kim:1979if} 
  J.~E.~Kim,
  Phys.\ Rev.\ Lett.\  {\bf 43}, 103 (1979).
  doi:10.1103/PhysRevLett.43.103

\bibitem{Shifman:1979if} 
  M.~A.~Shifman, A.~I.~Vainshtein and V.~I.~Zakharov,
  Nucl.\ Phys.\ B {\bf 166}, 493 (1980).
  doi:10.1016/0550-3213(80)90209-6

\bibitem{Lee:2016ejx} 
  H.~S.~Lee and M.~S.~Seo,
  Phys.\ Lett.\ B {\bf 767}, 69 (2017)
  doi:10.1016/j.physletb.2017.01.058
  [arXiv:1608.02708 [hep-ph]].
  
\bibitem{Jaeckel:2013ija} 
  J.~Jaeckel,
  Frascati Phys.\ Ser.\  {\bf 56}, 172 (2012)
  [arXiv:1303.1821 [hep-ph]].
  
\bibitem{Aubert:2001tu} 
  B.~Aubert {\it et al.} [BaBar Collaboration],
  Nucl.\ Instrum.\ Meth.\ A {\bf 479}, 1 (2002)
  doi:10.1016/S0168-9002(01)02012-5
  [hep-ex/0105044].

\bibitem{Lees:2013rw} 
  J.~P.~Lees {\it et al.} [BaBar Collaboration],
  Nucl.\ Instrum.\ Meth.\ A {\bf 726}, 203 (2013)
  doi:10.1016/j.nima.2013.04.029
  [arXiv:1301.2703 [hep-ex]].
  
\bibitem{Shtabovenko:2016sxi} 
  V.~Shtabovenko, R.~Mertig and F.~Orellana,
  Comput.\ Phys.\ Commun.\  {\bf 207}, 432 (2016)
  doi:10.1016/j.cpc.2016.06.008
  [arXiv:1601.01167 [hep-ph]].
  
\bibitem{Mertig:1990an}
  R.~Mertig, M.~Bohm and A.~Denner,
  Comput.\ Phys.\ Commun.\  {\bf 64}, 345 (1991).
  doi:10.1016/0010-4655(91)90130-D
  
  \bibitem{Essig:2013vha} 
  R.~Essig, J.~Mardon, M.~Papucci, T.~Volansky and Y.~M.~Zhong,
  JHEP {\bf 1311}, 167 (2013)
  doi:10.1007/JHEP11(2013)167
  [arXiv:1309.5084 [hep-ph]].
  
\bibitem{Aubert:2008as} 
  B.~Aubert {\it et al.} [BaBar Collaboration],
  arXiv:0808.0017 [hep-ex].
  
\bibitem{Lees:2017lec} 
  J.~P.~Lees {\it et al.} [BaBar Collaboration],
  Phys.\ Rev.\ Lett.\  {\bf 119}, no. 13, 131804 (2017)
  doi:10.1103/PhysRevLett.119.131804
  [arXiv:1702.03327 [hep-ex]].
  
\bibitem{Pukhov:1999gg} 
  A.~Pukhov {\it et al.},
  hep-ph/9908288.
  
\bibitem{Belyaev:2012qa} 
  A.~Belyaev, N.~D.~Christensen and A.~Pukhov,
  Comput.\ Phys.\ Commun.\  {\bf 184}, 1729 (2013)
  doi:10.1016/j.cpc.2013.01.014
  [arXiv:1207.6082 [hep-ph]].
  
\bibitem{Skwarnicki:1986xj} 
  T.~Skwarnicki,
  DESY-F31-86-02, DESY-F-31-86-02.

\bibitem{Abe:2010gxa} 
  T.~Abe {\it et al.} [Belle-II Collaboration],
  arXiv:1011.0352 [physics.ins-det].

\bibitem{HeartyBelleII}
\bibinfo{author}{C.~Hearty},
  \emph{\bibinfo{title}{{Dark sector searches at B-factories and outlook for
  Belle II}}},
  \bibinfo{howpublished}{\url{https://indico.fnal.gov/getFile.py/access?contribId=123&sessionId=9&resId=0&materialId=slides&confId=13702}}.

\bibitem{Jegerlehner:2009ry} 
  F.~Jegerlehner and A.~Nyffeler,
  Phys.\ Rept.\  {\bf 477}, 1 (2009)
  doi:10.1016/j.physrep.2009.04.003
  [arXiv:0902.3360 [hep-ph]].
  
\bibitem{Lindner:2016bgg} 
  M.~Lindner, M.~Platscher and F.~S.~Queiroz,
  Phys.\ Rept.\  {\bf 731}, 1 (2018)
  doi:10.1016/j.physrep.2017.12.001
  [arXiv:1610.06587 [hep-ph]].

\bibitem{Mohr:2012tt} 
  P.~J.~Mohr, B.~N.~Taylor and D.~B.~Newell,
  Rev.\ Mod.\ Phys.\  {\bf 84}, 1527 (2012)
  doi:10.1103/RevModPhys.84.1527
  [arXiv:1203.5425 [physics.atom-ph]].

\bibitem{Bennett:2002jb} 
  G.~W.~Bennett {\it et al.} [Muon g-2 Collaboration],
  Phys.\ Rev.\ Lett.\  {\bf 89}, 101804 (2002)
  Erratum: [Phys.\ Rev.\ Lett.\  {\bf 89}, 129903 (2002)]
  doi:10.1103/PhysRevLett.89.129903, 10.1103/PhysRevLett.89.101804
  [hep-ex/0208001].
   
\bibitem{Bennett:2004pv} 
  G.~W.~Bennett {\it et al.} [Muon g-2 Collaboration],
  Phys.\ Rev.\ Lett.\  {\bf 92}, 161802 (2004)
  doi:10.1103/PhysRevLett.92.161802
  [hep-ex/0401008].

\bibitem{Bennett:2006fi} 
  G.~W.~Bennett {\it et al.} [Muon g-2 Collaboration],
  Phys.\ Rev.\ D {\bf 73}, 072003 (2006)
  doi:10.1103/PhysRevD.73.072003
  [hep-ex/0602035].
  
\bibitem{Grange:2015fou} 
  J.~Grange {\it et al.} [Muon g-2 Collaboration],
  arXiv:1501.06858 [physics.ins-det].
  
\bibitem{Bouchendira:2010es} 
  R.~Bouchendira, P.~Clade, S.~Guellati-Khelifa, F.~Nez and F.~Biraben,
  Phys.\ Rev.\ Lett.\  {\bf 106}, 080801 (2011)
  doi:10.1103/PhysRevLett.106.080801
  [arXiv:1012.3627 [physics.atom-ph]].
  
\bibitem{Hanneke:2008tm}
  D.~Hanneke, S.~Fogwell and G.~Gabrielse,
  Phys.\ Rev.\ Lett.\  {\bf 100}, 120801 (2008)
  doi:10.1103/PhysRevLett.100.120801
  [arXiv:0801.1134 [physics.atom-ph]].
  
\bibitem{Hanneke:2010au}
  D.~Hanneke, S.~F.~Hoogerheide and G.~Gabrielse,
  Phys.\ Rev.\ A {\bf 83}, 052122 (2011)
  doi:10.1103/PhysRevA.83.052122
  [arXiv:1009.4831 [physics.atom-ph]].
  
\bibitem{Aoyama:2012wj} 
  T.~Aoyama, M.~Hayakawa, T.~Kinoshita and M.~Nio,
  Phys.\ Rev.\ Lett.\  {\bf 109}, 111807 (2012)
  doi:10.1103/PhysRevLett.109.111807
  [arXiv:1205.5368 [hep-ph]].

\bibitem{Endo:2012hp} 
  M.~Endo, K.~Hamaguchi and G.~Mishima,
  Phys.\ Rev.\ D {\bf 86}, 095029 (2012)
  doi:10.1103/PhysRevD.86.095029
  [arXiv:1209.2558 [hep-ph]].

\bibitem{Marciano:2016yhf} 
  W.~J.~Marciano, A.~Masiero, P.~Paradisi and M.~Passera,
  Phys.\ Rev.\ D {\bf 94}, no. 11, 115033 (2016)
  doi:10.1103/PhysRevD.94.115033
  [arXiv:1607.01022 [hep-ph]].
  
\bibitem{Pospelov:2007mp} 
  M.~Pospelov, A.~Ritz and M.~B.~Voloshin,
  Phys.\ Lett.\ B {\bf 662}, 53 (2008)
  doi:10.1016/j.physletb.2008.02.052
  [arXiv:0711.4866 [hep-ph]].
  
\bibitem{Batell:2009di} 
  B.~Batell, M.~Pospelov and A.~Ritz,
  Phys.\ Rev.\ D {\bf 80}, 095024 (2009)
  doi:10.1103/PhysRevD.80.095024
  [arXiv:0906.5614 [hep-ph]].

\bibitem{Jaeger:1974pk} 
  K.~Jaeger {\it et al.},
  Phys.\ Rev.\ D {\bf 11}, 1756 (1975).
  doi:10.1103/PhysRevD.11.1756
  
\bibitem{Amaldi:1979zk} 
  E.~Amaldi {\it et al.},
  Nucl.\ Phys.\ B {\bf 158}, 1 (1979).
  doi:10.1016/0550-3213(79)90183-4
  
\bibitem{deNiverville:2016rqh} 
  P.~deNiverville, C.~Y.~Chen, M.~Pospelov and A.~Ritz,
  Phys.\ Rev.\ D {\bf 95}, no. 3, 035006 (2017)
  doi:10.1103/PhysRevD.95.035006
  [arXiv:1609.01770 [hep-ph]].
  
\bibitem{Athanassopoulos:1996ds} 
  C.~Athanassopoulos {\it et al.} [LSND Collaboration],
  Nucl.\ Instrum.\ Meth.\ A {\bf 388}, 149 (1997)
  doi:10.1016/S0168-9002(96)01155-2
  [nucl-ex/9605002].
  
\bibitem{Aguilar:2001ty} 
  A.~Aguilar-Arevalo {\it et al.} [LSND Collaboration],
  Phys.\ Rev.\ D {\bf 64}, 112007 (2001)
  doi:10.1103/PhysRevD.64.112007
  [hep-ex/0104049].

\bibitem{deNiverville:2011it} 
  P.~deNiverville, M.~Pospelov and A.~Ritz,
  Phys.\ Rev.\ D {\bf 84}, 075020 (2011)
  doi:10.1103/PhysRevD.84.075020
  [arXiv:1107.4580 [hep-ph]].

\bibitem{Kahn:2014sra} 
  Y.~Kahn, G.~Krnjaic, J.~Thaler and M.~Toups,
  Phys.\ Rev.\ D {\bf 91}, no. 5, 055006 (2015)
  doi:10.1103/PhysRevD.91.055006
  [arXiv:1411.1055 [hep-ph]].

\bibitem{Burman:1989ds} 
  R.~L.~Burman and E.~S.~Smith,
  LA-11502-MS.
  
\bibitem{Auerbach:2001wg} 
  L.~B.~Auerbach {\it et al.} [LSND Collaboration],
  Phys.\ Rev.\ D {\bf 63}, 112001 (2001)
  doi:10.1103/PhysRevD.63.112001
  [hep-ex/0101039].
  
\bibitem{AguilarArevalo:2008yp} 
  A.~A.~Aguilar-Arevalo {\it et al.} [MiniBooNE Collaboration],
  Phys.\ Rev.\ D {\bf 79}, 072002 (2009)
  doi:10.1103/PhysRevD.79.072002
  [arXiv:0806.1449 [hep-ex]].
  
\bibitem{Aguilar-Arevalo:2012fmn} 
  A.~A.~Aguilar-Arevalo {\it et al.} [MiniBooNE Collaboration],
  arXiv:1207.4809 [hep-ex].
  
\bibitem{Aguilar-Arevalo:2013nkf} 
  A.~A.~Aguilar-Arevalo {\it et al.} [MiniBooNE Collaboration],
  Phys.\ Rev.\ D {\bf 91}, no. 1, 012004 (2015)
  doi:10.1103/PhysRevD.91.012004
  [arXiv:1309.7257 [hep-ex]].
  
\bibitem{Aguilar-Arevalo:2017mqx}
  A.~A.~Aguilar-Arevalo {\it et al.} [MiniBooNE Collaboration],
  Phys.\ Rev.\ Lett.\  {\bf 118}, no. 22, 221803 (2017)
  doi:10.1103/PhysRevLett.118.221803
  [arXiv:1702.02688 [hep-ex]].
  
\bibitem{Aguilar-Arevalo:2018wea} 
  A.~A.~Aguilar-Arevalo {\it et al.} [MiniBooNE DM Collaboration],
  arXiv:1807.06137 [hep-ex].

\bibitem{AguilarArevalo:2008qa} 
  A.~A.~Aguilar-Arevalo {\it et al.} [MiniBooNE Collaboration],
  Nucl.\ Instrum.\ Meth.\ A {\bf 599}, 28 (2009)
  doi:10.1016/j.nima.2008.10.028
  [arXiv:0806.4201 [hep-ex]].
  
\bibitem{deNiverville:2012ij} 
  P.~deNiverville, D.~McKeen and A.~Ritz,
  Phys.\ Rev.\ D {\bf 86}, 035022 (2012)
  doi:10.1103/PhysRevD.86.035022
  [arXiv:1205.3499 [hep-ph]].

\bibitem{Gninenko:2011uv} 
  S.~N.~Gninenko,
  Phys.\ Rev.\ D {\bf 85}, 055027 (2012)
  doi:10.1103/PhysRevD.85.055027
  [arXiv:1112.5438 [hep-ph]].
  
\bibitem{Gninenko:2012eq} 
  S.~N.~Gninenko,
  Phys.\ Lett.\ B {\bf 713}, 244 (2012)
  doi:10.1016/j.physletb.2012.06.002
  [arXiv:1204.3583 [hep-ph]].
  
\bibitem{Bonesini:2001iz} 
  M.~Bonesini, A.~Marchionni, F.~Pietropaolo and T.~Tabarelli de Fatis,
  Eur.\ Phys.\ J.\ C {\bf 20}, 13 (2001)
  doi:10.1007/s100520100656
  [hep-ph/0101163].
  
\bibitem{Dorenbosch:1987pn} 
  J.~Dorenbosch {\it et al.} [CHARM Collaboration],
  Z.\ Phys.\ C {\bf 40}, 497 (1988).
  doi:10.1007/BF01560221
  
\bibitem{Bergsma:1983rt} 
  F.~Bergsma {\it et al.} [CHARM Collaboration],
  Phys.\ Lett.\  {\bf 128B}, 361 (1983).
  doi:10.1016/0370-2693(83)90275-7
  
\bibitem{Bergsma:1985is} 
  F.~Bergsma {\it et al.} [CHARM Collaboration],
  Phys.\ Lett.\  {\bf 166B}, 473 (1986).
  doi:10.1016/0370-2693(86)91601-1
  
\bibitem{Bjorken:1988as} 
  J.~D.~Bjorken {\it et al.},
  Phys.\ Rev.\ D {\bf 38}, 3375 (1988).
  doi:10.1103/PhysRevD.38.3375
  
\bibitem{Batell:2014mga} 
  B.~Batell, R.~Essig and Z.~Surujon,
  Phys.\ Rev.\ Lett.\  {\bf 113}, no. 17, 171802 (2014)
  doi:10.1103/PhysRevLett.113.171802
  [arXiv:1406.2698 [hep-ph]].
  
\bibitem{Lai:2010vv} 
  H.~L.~Lai, M.~Guzzi, J.~Huston, Z.~Li, P.~M.~Nadolsky, J.~Pumplin and C.-P.~Yuan,
  Phys.\ Rev.\ D {\bf 82}, 074024 (2010)
  doi:10.1103/PhysRevD.82.074024
  [arXiv:1007.2241 [hep-ph]].

\bibitem{Atiya:1992sm} 
  M.~S.~Atiya {\it et al.},
  Phys.\ Rev.\ Lett.\  {\bf 69}, 733 (1992).
  doi:10.1103/PhysRevLett.69.733
  
\bibitem{Artamonov:2008qb} 
  A.~V.~Artamonov {\it et al.} [E949 Collaboration],
  Phys.\ Rev.\ Lett.\  {\bf 101}, 191802 (2008)
  doi:10.1103/PhysRevLett.101.191802
  [arXiv:0808.2459 [hep-ex]].

  A.~V.~Artamonov {\it et al.} [BNL-E949 Collaboration],
  Phys.\ Rev.\ D {\bf 79}, 092004 (2009)
  doi:10.1103/PhysRevD.79.092004
  [arXiv:0903.0030 [hep-ex]].
  
\bibitem{Achasov:2000zd} 
  M.~N.~Achasov {\it et al.},
  Eur.\ Phys.\ J.\ C {\bf 12}, 25 (2000).
  doi:10.1007/s100529900222
  
\bibitem{Anelli:2015pba} 
  M.~Anelli {\it et al.} [SHiP Collaboration],
  arXiv:1504.04956 [physics.ins-det].

\bibitem{Tufanli:2016hyo} 
  S.~Tufanli [SBND Collaboration],
  PoS HQL {\bf 2016}, 070 (2017).
  doi:10.22323/1.274.0070
  
\bibitem{Ko:2016owz} 
  Y.~J.~Ko {\it et al.},
  Phys.\ Rev.\ Lett.\  {\bf 118}, no. 12, 121802 (2017)
  doi:10.1103/PhysRevLett.118.121802
  [arXiv:1610.05134 [hep-ex]].

\bibitem{Park:2017prx} 
  H.~Park,
  Phys.\ Rev.\ Lett.\  {\bf 119}, no. 8, 081801 (2017)
  doi:10.1103/PhysRevLett.119.081801
  [arXiv:1705.02470 [hep-ph]].
  
\bibitem{Banerjee:2017hhz} 
  D.~Banerjee {\it et al.} [NA64 Collaboration],
  Phys.\ Rev.\ D {\bf 97}, no. 7, 072002 (2018)
  doi:10.1103/PhysRevD.97.072002
  [arXiv:1710.00971 [hep-ex]].
  
\bibitem{Dreiner:2013mua} 
  H.~K.~Dreiner, J.~F.~Fortin, C.~Hanhart and L.~Ubaldi,
  Phys.\ Rev.\ D {\bf 89}, no. 10, 105015 (2014)
  doi:10.1103/PhysRevD.89.105015
  [arXiv:1310.3826 [hep-ph]].

\bibitem{Redondo:2013lna} 
  J.~Redondo and G.~Raffelt,
  JCAP {\bf 1308}, 034 (2013)
  doi:10.1088/1475-7516/2013/08/034
  [arXiv:1305.2920 [hep-ph]].

\bibitem{Dolan:2017osp} 
  M.~J.~Dolan, T.~Ferber, C.~Hearty, F.~Kahlhoefer and K.~Schmidt-Hoberg,
  JHEP {\bf 1712}, 094 (2017)
  doi:10.1007/JHEP12(2017)094
  [arXiv:1709.00009 [hep-ph]].
  
\bibitem{Lee:1969fy} 
  T.~D.~Lee and G.~C.~Wick,
  Nucl.\ Phys.\ B {\bf 9}, 209 (1969).
  doi:10.1016/0550-3213(69)90098-4

\bibitem{Lee:1970iw} 
  T.~D.~Lee and G.~C.~Wick,
  Phys.\ Rev.\ D {\bf 2}, 1033 (1970).
  doi:10.1103/PhysRevD.2.1033

\bibitem{Grinstein:2007mp} 
  B.~Grinstein, D.~O'Connell and M.~B.~Wise,
  Phys.\ Rev.\ D {\bf 77}, 025012 (2008)
  doi:10.1103/PhysRevD.77.025012
  [arXiv:0704.1845 [hep-ph]].

\bibitem{Cutkosky:1969fq} 
  R.~E.~Cutkosky, P.~V.~Landshoff, D.~I.~Olive and J.~C.~Polkinghorne,
  Nucl.\ Phys.\ B {\bf 12}, 281 (1969).
  doi:10.1016/0550-3213(69)90169-2

\bibitem{Lee:1971ix} 
  T.~D.~Lee and G.~C.~Wick,
  Phys.\ Rev.\ D {\bf 3}, 1046 (1971).
  doi:10.1103/PhysRevD.3.1046
    
 \bibitem{Coleman:1970lec}
 S. Coleman, ``Acausality", in “Erice 1969, Ettore Majorana School On Subnuclear Phenomena”, New York, 282 (1970).
 
  
\end{thebibliography}
\end{document}